\documentclass[aps,pre,amsmath,amssymb,eqsecnum,amsfonts, reprint,%
amssymb,showkeys,showpacs,bm]{revtex4-1}

\usepackage{graphicx}

\begin{document}

\title{Anomalous scaling of passive scalar fields \\
advected by the Navier-Stokes velocity ensemble: \\
Effects of strong compressibility and large-scale anisotropy}

\author{N.\,V.~Antonov and M.\,M.~Kostenko}

\email{n.antonov@spbu.ru, kontramot@mail.ru}

\affiliation{Chair of High Energy Physics and Elementary Particles \\
Department of Theoretical Physics, Faculty of Physics \\
Saint Petersburg State University, Ulyanovskaja~1 \\
Saint~Petersburg--Petrodvorez, 198904 Russia}

\begin{abstract}
The field theoretic renormalization group and the operator product
expansion are applied to two models of passive scalar quantities
(the density and the tracer fields) advected by a  random turbulent
velocity field. The latter is governed by the Navier--Stokes equation
for compressible fluid, subject to external random force with the
covariance $\propto \delta(t-t') k^{4-d-y}$, where $d$ is the dimension
of space and $y$ is an arbitrary exponent. The original stochastic problems
are reformulated as multiplicatively renormalizable field theoretic models;
the corresponding renormalization group equations possess infrared
attractive fixed points. It is shown that various correlation functions
of the scalar field, its powers and gradients, demonstrate anomalous
scaling behavior in the inertial-convective range already for small values
of $y$. The corresponding anomalous exponents, identified with scaling
(critical) dimensions of certain composite fields (``operators'' in the
quantum-field terminology), can be systematically calculated as series
in $y$. The practical calculation is performed in the leading one-loop
approximation, including exponents in anisotropic contributions.
It should be emphasized that, in contrast to Gaussian ensembles with
finite correlation time, the model and the perturbation theory presented
here are manifestly Galilean covariant. The validity of the one-loop
approximation and comparison with Gaussian models are briefly discussed.
\end{abstract}

\pacs{47.27.eb, 47.27.ef, 05.10.Cc}

\keywords{fully developed turbulence, passive advection, anomalous scaling,
renormalization group, operator product expansion, composite fields,
compressibility, anisotropy}


\maketitle

\section{Introduction} \label{sec:Intro}

In a few past decades, intermittent interest has been attracted to the
problem of intermittency and anomalous scaling in fluid turbulence; see
{\it e.g.} Refs. \cite{Legacy}--\cite{GK} and the literature cited therein.
The phenomenon manifests itself in singular (arguably power-like) behavior
of various statistical quantities as functions of the integral turbulence
scales, with infinite sets of independent anomalous exponents \cite{Legacy}.
In spite of considerable success, the problem remains essentially open: no
regular calculational scheme, based on an underlying dynamical model and
reliable perturbation expansion (like the famous $\varepsilon$ expansion for
critical exponents) was ever constructed for the anomalous exponents of the
turbulent velocity field.

Both the natural experiments and numerical simulations suggest that
deviations from the classical Kolmogorov theory are even more strongly
pronounced for passively advected scalar fields (like the temperature or
the density of a pollutant) than for the velocity field itself
\cite{NonG}--\cite{ShS}. At the same time, various simplified models,
describing passive advection by  ``synthetic'' velocity fields with given
statistics, appear easier tractable theoretically and allow analytical
results to be derived \cite{FGV}. Therefore, the problem of passive
advection, being of practical importance in itself, may also be viewed
as a starting point in studying intermittency and anomalous scaling in
fluid turbulence on the whole.

The most remarkable progress was achieved for the Kraichnan's
rapid-change model \cite{Kraich2}, where the advecting velocity field is
taken Gaussian, not correlated in time, and having a power-like correlation
function of the form $\sim\delta(t-t') /k^{d+\xi}$, where $d$ is the
dimension of space, $k$ is the wave number and $\xi$ is an arbitrary
exponent. There, for the first time, the existence of anomalous scaling
was firmly established on the
basis of a microscopic model \cite{Kraich2}; the corresponding anomalous
exponents were calculated in controlled approximations \cite{GK} and,
eventually, within a systematic perturbation expansion in a formal small
parameter $\xi$ \cite{RG}. Detailed review of the theoretical research on the
passive scalar problem and the bibliography can be found in Ref.~\cite{FGV}.

In the original Kraichnan's model, the velocity ensemble was taken Gaussian,
decorrelated in time, isotropic, and the fluid was implied to be
incompressible.
More realistic models should take into account finite correlation time and
non-Gaussianity of the velocity ensemble, anisotropy of the experimental
set-up, compressibility of the fluid, {\it etc.}; see the discussion in
\cite{NonG,OU}. Here, the two key issue arise:
formulation of more realistic models and the possibility to
treat them (more or less) analytically.

A most efficient way to study anomalous scaling is provided by the field
theoretic renormalization group (RG) combined with the operator product
expansion (OPE); see the books \cite{Zinn,Book3} for the detailed
exposition of these techniques and the references. In the RG+OPE scenario
for the anomalous scaling in turbulence, proposed in \cite{JETP},
the singular dependence on the integral scales
emerges as a consequence of the existence in the corresponding models of
composite fields (``composite operators'' in the quantum-field terminology)
with {\it negative} scaling dimensions; termed as ``dangerous operators;''
for more detailed explanations and the references, see
\cite{Book3,JETP,UFN,turbo}. For Kraichnan's model, the anomalous exponents
can be identified with the scaling dimensions (``critical dimensions''
in the terminology of the theory of critical state) of certain individual
Galilean-invariant composite operators \cite{RG}. This allows one to give a
self-consistent derivation of the anomalous scaling, to construct a
systematic perturbation expansion for the anomalous exponents in $\xi$, and
to calculate the exponents in the second \cite{RG} and in the third
\cite{cube} orders. The RG approach can be generalized to the case of finite
correlation
time \cite{A99} and to the non-Gaussian advecting velocity field, governed
by the stochastic Navier--Stokes equation \cite{NSpass}. A general
overview of the RG approach to Kraichnan's model and its descendants
and more references can be found in \cite{JphysA}.

Numerous studies were devoted to the effects of compressibility on the
intermittency and anomalous scaling \cite{El}--\cite{AK2}. Analysis of
simplified models suggests that compressibility strongly affects the
passively advected fields. In particular, in contrast to the incompressible
case, the diffusion can be depleted by the advection of a purely potential
flow \cite{Avell} and the phase transition from a turbulent to a certain
purely chaotic state takes place
when the degree of compressibility increases \cite{tracer}. It was also
shown that the anomalous exponents become non-universal due to dependence
on the compressibility parameter, such that the anomalous scaling is
enhanced, while the hierarchy of anisotropic contributions is
suppressed \cite{RG1}--\cite{AK2}. For passive vector ({\it e.g.} magnetic)
fields, the issues of anomalous scaling, hierarchy of anisotropic
contributions and the dependence on compressibility were discussed
{\it e.g.} in \cite{V96}--\cite{J13}.

An important advantage of Kraichnan's model is the possibility to easily
model compressibility \cite{El}--\cite{RG1}. Generalization to the case
of a Gaussian ensemble with finite correlation time is also possible
\cite{AKens,AK2,TWP}. However, synthetic models with non-vanishing
correlation time suffer from the lack of Galilean symmetry, which may
lead to ``interesting pathologies'' (quoting Ref.~\cite{OU}). In the RG
approach, one of such a pathology manifests itself as an ultraviolet (UV)
divergence in the vertex \cite{AKens}, which in more realistic models is
forbidden by Galilean invariance, and for the incompressible Gaussian
model is absent because of rather technical reasons \cite{A99}.
Thus it is desirable to describe the advecting velocity field by the
corresponding Navier--Stokes equations \cite{LL} with a random stirring
force. However, this appeared to be a difficult task.

In Refs. \cite{Lviv,KompOp}, the leading-order correction in the Mach number
Ma to the incompressible scaling regime was studied; generalization to all
orders of the expansion in Ma was derived in \cite{NV}. The corrections are
small for very small Ma and not very small momenta $k$, but become
arbitrarily
large (IR relevant in the sense of Wilson) and destroy the incompressible
scaling regime if Ma is fixed and the momenta become small enough. Thus the
original incompressible regime becomes unstable, and a crossover to another
unknown regime occurs. The case of strong compressibility was studied in
Refs.~\cite{Tur}--\cite{ANU}. The results are rather controversial, but all
of those studies support the existence of a stationary resulting
``compressible'' regime, different from the original incompressible one.

In the present paper, we adopt the approach of Ref.~\cite{ANU}, where the
standard field theoretic RG was applied to the problem of stirred
hydrodynamics of a compressible fluid, and the resulting stationary scaling
regime was associated with the IR attractive fixed point of the corresponding
multiplicatively renormalizable field theoretic model. That approach was
later applied to the problem of mass distribution in the self-gravitating
matter within the framework of a continuous stochastic formulation of the
Vlasov--Poisson model \cite{PRL}. The problem of anomalous scaling of the
velocity field in that model remains open, as for its incompressible
predecessors, but the passive scalar advection by such an ensemble can be
treated analytically. This is the aim of the present work.

The plan of the paper is as follows.

In section~\ref{sec:NScomp} we revisit the RG approach to the stochastic
Navier--Stokes equation for a compressible fluid, following mostly
Ref.~\cite{ANU},
and introduce the basic notions (field theoretic formulation, canonical
dimensions, renormalizability and RG equations), needed for the further
analysis of the passive advection. The RG equations possess an IR
attractive fixed point, which implies existence of a scaling regime in
the inertial and energy-containing ranges. The one-loop explicit expressions
for the renormalization constants and the RG functions (anomalous dimensions
and $\beta$ functions), calculated in \cite{ANU}, are presented.
The corresponding scaling dimensions of the frequency and the velocity
are known exactly and coincide with their analogs for the incompressible
case. Another nontrivial fixed point is unstable (it is a saddle point) and
corresponds to the incompressible fluid.

In section~\ref{sec:PD} we introduce the diffusion-advection stochastic
equations for the two types of passive scalar field: the tracer (temperature,
entropy or concentration of a pollutant) and the density of a conserved
quantity ({\it e.g.} density of a pollutant). We present the field theoretic
formulation of these models and show that they are multiplicatively
renormalizable. Then the RG equations can be derived in a standard fashion.
The renormalization constants and the RG functions are calculated in the
leading (one-loop) approximation, which is consistent with the accuracy of
the results derived in \cite{ANU}. The full-scale models, involving the
velocity field and the scalar field, possess an IR attractive fixed point.
Thus the existence of a scaling regime in the IR range is established. Exact
expressions for the scaling dimensions of the scalar fields are obtained.

In section~\ref{sec:Opera} we calculate, in the leading order of the
expansion in $y$ (one-loop approximation), critical dimensions of the
composite operators built of the scalar field and its spatial derivatives,
including some tensor operators. In the next section, those dimensions
are identified with various anomalous exponents.

In section~\ref{sec:OPE} we apply the OPE to the analysis of the inertial-range
behavior of various correlation functions: the correlation functions of
the scalar fields and their powers for the density case and of the structure
functions for the tracer case. We show that, for the density case,
leading terms of the inertial-range behavior are determined by the
contributions of the operators built solely of the scalar fields.
Their critical dimensions are negative, which leads to strong dependence
on the integral scale and to the anomalous scaling, with the anomalous
exponents identified with those dimensions.

For the tracer case, more interesting quantities are the structure
functions that involve differences of the values of the scalar field
at different points. Their anomalous behaviour is determined by the
scalar operators built of the gradients of the scalar field, whose
negative dimensions are identified with the corresponding anomalous
exponents.

In the presence of anisotropy, introduced into the system at large scales,
contributions of the tensor operators in the OPE's come into play: $l$th
rank tensor operators determine the contribution in the correlation functions
with nontrivial angular dependence described by the $l$th order spherical
harmonics. Like for the Kraichnan model, those anisotropic contributions
organize a kind of hierarchy, related to the degree of anisotropy: they
become less important as $l$ grows, so that the leading term of the
inertial-range asymptotic behavior is given by the isotropic contribution
($l=0$) in agreement with Kolmogorov's hypothesis of the local isotropy
restoration. This issue is discussed for the pair correlation function in
the both models and for the structure functions of arbitrary order for the
tracer.

Section \ref{sec:DC14} is reserved for the discussion, comparison with the
Gaussian models and the conclusion.

\section{RG analysis of the stochastic NS equation with strong
compressibility} \label{sec:NScomp}

\subsection{Description of the model}
\label{sec:NScomp1}

The Navier--Stokes equation for a viscid compressible fluid has the
form \cite{LL}
\begin{eqnarray}
\rho\nabla_{t} v_{i} = \nu_{0} [\delta_{ik}\partial^{2}-
\partial_{i}\partial_{k}] v_{k}
+ \mu_0 \partial_{i}\partial_{k} v_{k} - \partial_{i} p + \eta_{i},
\nonumber \\ {}
\label{NS}
\end{eqnarray}
where
\begin{equation}
\nabla_{t} = \partial_{t} + v_{k} \partial_{k}
\label{nabla}
\end{equation}
is the Lagrangian (Galilean covariant) derivative, $\partial_{t} =
\partial /\partial t$, $\partial_{i} = \partial /\partial x_{i}$, and
$\partial^{2} =\partial_{i}\partial_{i}$ is the Laplace operator.

Equation (\ref{NS}) is obtained by combining the momentum balance equation
\begin{eqnarray}
\partial_{t} (\rho v_{i}) + \partial_{k} \Pi_{ik} = \eta_{i},
\label{MB}
\end{eqnarray}
where
\begin{eqnarray}
\Pi_{ik} = \rho v_{i}v_{k}+\delta_{ik}p -\nu_0 (\partial_{i}v_{k}+
\partial_{k}v_{i}\!) - \delta_{ik} (\mu_0-2\nu_0\!)\, \partial_{l}v_{l}
\nonumber \\ {}
\label{ST}
\end{eqnarray}
is the stress tensor, with the continuity equation
\begin{eqnarray}
\partial_{t} \rho  + \partial_{i} (\rho v_{i})   = 0.
\label{CE}
\end{eqnarray}

In those equations, $v_{i}$ is the velocity, $\rho$ is the mass density,
$p$ is the pressure, and $\eta_{i}$ is the density of the external force
(per unit volume). All these quantities depend on $x=\{t,{\bf x}\}$ with
${\bf x}=\{x_i\}$, $i=1\dots d$, where $d$ is an arbitrary (for generality)
dimensionality of space. The constants $\nu_{0}$ and $\mu_{0}$ are two
independent molecular viscosity coefficients; in the viscous terms in
(\ref{NS}) we explicitly separated the transverse and the longitudinal
parts. Summations over repeated vector indices are always implied.

The problem (\ref{NS}), (\ref{CE}) should be augmented by the equation of
state, $p=p(\rho)$. It will be taken in the simplest form of the linear
relation
\begin{eqnarray}
(p-\bar p) = c^{2}_{0} (\rho-\bar\rho)
\label{ES}
\end{eqnarray}
between the deviations of the pressure and the density from their mean
values. The constant $c_{0}$ has the meaning of the (adiabatic) speed of
sound.

Following \cite{ANU}, we divide equation (\ref{NS}) with $\rho$ and in the
viscous terms replace $\rho$ with its mean value. This approximation
(which is needed to obtain a renormalizable local field theoretic model)
is implicitly justified by the analysis of Ref. \cite{NV}; we also note that
the viscosity plays a little role in the inertial range. We retain the same
notation $\nu_{0}$ and $\mu_{0}$ for the resulting constant kinematic
viscosity coefficients. Then the equations (\ref{NS}), (\ref{CE}) can
be rewritten in the form
\begin{eqnarray}
\nabla_{t} v_{i} &=&
\nu_{0} [\delta_{ik}\partial^{2}-\partial_{i}\partial_{k}]
v_{k}\! +\! \mu_0 \partial_{i}\partial_{k} v_{k} -\!
\partial_{i} \phi\! +\! f_{i},
\label{ANU} \\
\nabla_{t} \phi &=& -c_{0}^{2} \partial_{i}v_{i},
\label{ANU1}
\end{eqnarray}
where we have introduced the new scalar field
\begin{eqnarray}
\phi = c_{0}^{2} \ln (\rho/\bar \rho)
\label{fi}
\end{eqnarray}
and $f_{i}=f_{i}(x)$ is the density of the external force (per unit mass).

In the stochastic formulation of the problem, the external force becomes a
random field that models the energy input into the system from the
large-scale stirring. The details of its statistics are believed to be
unessential, so it is taken to be Gaussian with zero mean, white in-time
(this is required by the Galilean symmetry), and involving some typical IR
scale $L$ (the integral scale). On the other hand, for the use of the
standard RG technique it is important that its correlation function have
a power-law tail at large wave numbers. More detailed
discussion can be found in \cite{UFN,turbo,China}. In the present case
one choses the correlation function in the form \cite{ANU}
\begin{eqnarray}
\langle f_{i}(x) f_{j}(x') \rangle = \delta(t-t') \int_{k>m} \frac{d{\bf k}}
{(2\pi)^{d}} \, D^{f}_{ij}({\bf k}) \exp\{{\rm i} {\bf kx}\},
\nonumber \\ {}
\label{force}
\end{eqnarray}
where
\begin{eqnarray}
D^{f}_{ij}({\bf k}) = D_{0}\, k^{4-d-y}\,
\left\{ P^{\bot}_{ij} ({\bf k})
+ \alpha P^{\parallel}_{ij} ({\bf k}) \right\}.
\label{power}
\end{eqnarray}
Here $P^{\bot}_{ij} ({\bf k})=\delta_{ij}-k_{i}k_{j}/k^{2}$ and
$P^{\parallel}_{ij} ({\bf k})=k_{i}k_{j}/k^{2}$ are the transverse and the
longitudinal projectors, respectively, $k=|{\bf k}|$ is the wave number,
$D_{0}$ and $\alpha$ are positive amplitudes. It is convenient to write
$D_{0}=g_{0}\nu_0^{3}$: the parameter $g_{0}$ plays the role of the coupling
constant (formal expansion parameter in the ordinary perturbation theory).
The relation $g_{0} \sim \Lambda^{y}$ sets in the typical UV momentum scale
(reciprocal of the dissipation length scale). The parameter $m = L^{-1}$
provides IR regularization; its precise form is unessential and the sharp
cut-off is the simplest choice for calculational reasons. The exponent
$0<y\le 4$ plays the role analogous to that played by $\varepsilon=4-d$
in the RG theory of critical behavior \cite{Zinn,Book3}: it provides UV
regularization (so that the UV divergences have the form of the poles in
$y$) and the
coordinates of fixed points and various scaling dimensions are calculated
as series in $y$. The most realistic (physical) value is given by the limit
$y\to4$, when the functions in (\ref{power}) can be viewed (with the proper
choice of the amplitude) as power-like models of the function
$\delta({\bf k})$: it corresponds to the idealized picture of the energy
input from infinitely large scales.

\subsection{Field theoretic formulation and Feynman rules}
\label{sec:NScomp2}

According to the general theorem \cite{Zinn,Book3}, the stochastic problem
(\ref{ANU}), (\ref{ANU1}), (\ref{force}), (\ref{power}),
is equivalent to the field theoretic model of the doubled set of fields
$\Phi=\left\{v_{i}',\phi', v_{i}, \phi\right\}$
and the action functional
\begin{widetext}
\begin{eqnarray}
{\cal S}(\Phi) &=& \frac{1}{2} v_{i}'D^{f}_{ik}  v_{k}' +
v_{i}' \left\{ -\nabla_{t} v_{i} +
\nu_{0} [\delta_{ik}\partial^{2}-\partial_{i}\partial_{k}] v_{k} +
u_0 \nu_{0}\partial_{i}\partial_{k} v_{k} - \partial_{i} \phi \right\} +
\nonumber \\
&+& \phi' \left[ -\nabla_{t} \phi  + v_0 \nu_{0} \partial^{2} \phi
- c_{0}^{2} (\partial_{i}v_{i}) \right],
\label{action}
\end{eqnarray}
\end{widetext}
where $D^{f}$ is the correlation function (\ref{force}), (\ref{power}),
and all the needed summations over the vector indices and integrations over
$x=\left\{t,{\bf x}\right\}$ are implied, for example,
\begin{eqnarray}
v_{i}' \nabla_{t} v_{i} = \int\! dt\! \int\! d{\bf x}\, v_{i}'(x)
[\partial_{t}+ v_{k}(x)\partial_{k}] v_{i}(x).
\label{rashif}
\end{eqnarray}
In (\ref{action}) we passed to the new dimensionless parameter
$u_0=\mu_0/\nu_0>0$ and introduced a new term
$v_0 \nu_{0} \phi' \partial^{2}\phi$ with another positive
dimensionless parameter $v_0$. This term is not forbidden by the symmetry or
dimensionality considerations, so it will necessarily appear in the
renormalization procedure. From the physics viewpoints, it corresponds to
some redefinition of the relation between the velocity and momentum \cite{LL}.
From a more technical point of view, it is needed to insure multiplicative
renormalizability of the model (\ref{action}), which allows one to easily
derive the RG equations. One can insist on studying the original model
(\ref{ANU}), (\ref{ANU1}) without such a term. Then the RG equations must
be solved with the initial condition $v_0=0$. In renormalized variables,
this corresponds to a general situation with a nonzero value of the
corresponding renormalized parameter $v$. Since the IR attractive fixed
point is unique (see below), the specific initial condition is unessential.

The field theoretic formulation means that various correlation functions
and response (Green) functions of the original stochastic problem are
represented by functional averages over the full set of fields with weight
$\exp {\cal S}(\Phi)$, and in this sense they can be viewed as the Green
functions of the field theoretic model with action (\ref{action}). The
model corresponds to standard Feynman diagrammatic techniques with two
vertices $-v'(v\partial)v$ and $-\phi'(v\partial)\phi$ and the free (bare)
propagators, determined by the quadratic part of the action functional; in
the frequency--momentum ($\omega$--${\bf k}$) representation, they have the
forms:
\begin{eqnarray}
\langle vv' \rangle_{0} &=& \langle v'v \rangle_{0}^{*}= P^{\bot}
\epsilon_{1}^{-1} + P^{\parallel} \epsilon_{3} R^{-1} ,
\nonumber \\
\langle vv \rangle_{0} &=&  P^{\bot} \frac{d^{f}}{|\epsilon_{1}|^{2}}
+ P^{\parallel}\alpha d^{f} \left|\frac{\epsilon_{3}}{R}\right|^{2},
\nonumber \\
\langle \phi v' \rangle_{0} &=& \langle v' \phi \rangle_{0}^{*}= -
\frac{{\rm i}c_{0}^{2}{\bf k}}{R}, \quad
\langle  v\phi' \rangle_{0} = \langle \phi'v \rangle_{0}^{*}=
\frac{{\rm i}{\bf k}}{R},
\nonumber \\
\langle  \phi\phi' \rangle_{0} &=& \langle \phi'\phi \rangle_{0}^{*}=
\frac{\epsilon_{2}}{R}, \quad
\langle  \phi\phi \rangle_{0} = \frac {\alpha c_{0}^{4} k^{2}d^{f}}
{|R|^{2}},
\nonumber \\
\langle  v\phi \rangle_{0} &=& \langle  \phi v \rangle_{0}^{*} =
\frac{ {\rm i}\alpha c_{0}^{2} d^{f}\epsilon_{3} {\bf k}} {|R|^{2}} ,
\nonumber \\
\langle  \phi'\phi' \rangle_{0} &=& \langle  v'\phi' \rangle_{0} =
\langle  v'v' \rangle_{0} = 0,
\label{lines}
\end{eqnarray}
where we have denoted
\begin{eqnarray}
\epsilon_{1}  &=& -{\rm i}\omega +\nu_0 k^{2}, \quad
\epsilon_{2}=-{\rm i}\omega+ u_0\nu_0 k^{2} ,
\nonumber \\
\epsilon_{3}  &=& -{\rm i}\omega+ v_0\nu_0 k^{2} , \quad
R=\epsilon_{2}\epsilon_{3}+c_{0}^{2}k^{2} ,
\nonumber \\
d^{f} &=& g_{0}\nu_0^{3}\, k^{4-d-y}
\label{energies}
\end{eqnarray}
and omitted the vector indices of the fields and the projectors.

In the limit $c_{0}\to\infty$, the propagators $\langle vv' \rangle_{0}$
and $\langle vv \rangle_{0}$ become purely transverse, while the mixed
propagator $\langle  v\phi \rangle_{0}$ vanishes. Then the scalar field
$\phi$ decouples from $v,v'$ (it does not enter the vertex in (\ref{ANU})),
and we arrive at the well-known Feynman rules for the incompressible
fluid \cite{Book3,UFN,turbo}.

\subsection{UV divergences, renormalization, and multiplicative
renormalizability} \label{sec:NScomp3}

It is well known that the analysis of UV divergences is based on the
analysis of canonical dimensions; see {\it e.g.}~\cite{Zinn,Book3}.
Dynamical models like (\ref{action}) have
two independent scales: the time scale $T$ and the length scale $L$. Thus
the canonical dimension of any quantity $F$ (a field or a parameter) is
described by two numbers, the
frequency dimension $d_{F}^{\omega}$ and the momentum dimension $d_{F}^{k}$,
defined such that $[F] \sim [T]^{-d_{F}^{\omega}} [L]^{-d_{F}^{k}}$.
The obvious consequences of the definition are the relations
\begin{eqnarray}
d_k^k=-d_{\bf x}^k=1,\ d_k^{\omega} =d_{\bf x}^{\omega }=0,
\nonumber \\
d_{\omega }^k=d_t^k=0, \ d_{\omega }^{\omega }=-d_t^{\omega }=1.
\label{canon}
\end{eqnarray}
The other dimensions are found from the requirement that each term of the
action functional be dimensionless (with respect to the momentum and the
frequency dimensions separately). Then one introduces the total canonical
dimension
\begin{eqnarray}
d_{F}=d_{F}^{k}+2d_{F}^{\omega},
\label{total}
\end{eqnarray}
which plays in the theory of renormalization of dynamical models the same
part as the conventional canonical dimension does in static problems.
The canonical dimensions for the model (\ref{action}) are given in
table~1, including renormalized parameters (without the subscript ``o''),
which will appear a bit later.

\begin{table*}
\caption{Canonical dimensions of the fields and parameters in the
models (\protect\ref{action}), (\protect\ref{Fact}),
(\protect\ref{Dact}), (\protect\ref{Tact}).}
\label{table1}
\begin{ruledtabular}
\begin{tabular}{c|c|c|c|c|c|c|c|c|c|c|c}
$F$ & $v'$ & $v$ & $\phi'$ & $\phi$ & $ \theta' $ &  $\theta$ &
$m$, $\mu$, $\Lambda$ & $\nu_0$, $\nu$  & $c_{0}$, $c$ &
$g_{0}$ & $u_{0}$, $v_{0}$ $w_{0}$, $u$, $v$, $w$, $g$, $\alpha$  \\
\hline
$d_{F}^{k}$ & $d+1$ & $-1$ & $d+2$ & $-2$ & $d$  & 0
& 1 &  $-2$ & $-1$ & $y$ & 0 \\
$d_{F}^{\omega}$ & $-1$ & 1 & $-2$ & 2 & $1/2$ & $-1/2$ &
0 & 1 & 1 & 0 & 0\\
$d_{F}$ & $d-1$ & 1 & $d-2$ & 2 & $d+1$ & $-1$ & 1 & 0 & 1 & $y$ & 0 \\
\end{tabular}
\end{ruledtabular}
\end{table*}

The choice (\ref{total}) for the total canonical dimension deserves
a more careful explanation. It means that all the viscosity or diffusion
coefficients in the model are pronounced dimensionless (with respect to
the new total dimension), and the time and the space variables are
measured in the same units; cf. \cite{Zinn,Book3}. Experienced reader
recalls the $c=1$ system of units in relativistic physics, where all the
distances are measured in the time units (light years).
Here, we relate the dimensions by eq. (\ref{total})
because the dispersion law for diffusion modes is $\omega \sim k^{2}$.
However, our model involves another dispersion law, $\omega \sim k$,
related to the sound modes. If we decide to set the speed of sound $c_{0}$
dimensionless, we would have to set $d_{F}=d_{F}^{k}+d_{F}^{\omega}$.

A similar alternative exists in the so-called model H of equilibrium
dynamical critical behavior, where the motion of the fluid is taken into
account and several dispersion laws are simultaneously present; see
{\it e.g.} p. 552 in the monograph \cite{Book3}. The choice
(\ref{total}) means that we are interested in the asymptotic behavior of
the Green functions where $\omega \sim k^{2} \to 0$; the RG treatment will
modify it to the Kolmogorov law $\omega \sim k^{2/3} \to 0$ (see below).
The same choice is made in the models of incompressible fluid (where
it is the only possible one because the speed of sound is infinite).
The alternative choice $d_{F}=d_{F}^{k}+d_{F}^{\omega}$ would mean that
we were interested in the asymptotic behavior of the (same) Green functions
for $\omega \sim k \to 0$ (sound modes in turbulent medium); this problem
is of course extremely interesting but so far it is not accessible by the
RG treatment, and will not be discussed in the present paper.

From table~1 it follows that the model becomes logarithmic (the coupling
constant $g_{0} \sim \Lambda^{y}$ becomes dimensionless) at $y=0$, so that
the UV divergences have the form of poles in $y$ in the Green functions.
The total canonical dimension of any 1-irreducible Green function $\Gamma$
(the formal index of UV divergence) is
\begin{eqnarray}
\delta_{\Gamma} = d+2 - \sum_{\Phi} N_{\Phi} d_{\Phi},
\label{index}
\end{eqnarray}
where $N_{\Phi}$ are the numbers of the fields entering into the function
$\Gamma$, $d_{\Phi}$ are their total canonical dimensions, and the summation
over all types of the fields $\Phi$ is implied. Superficial UV
divergences, whose removal requires counterterms, can be present only in
the functions $\Gamma$ with a non-negative integer $\delta_{\Gamma}$.
The counterterm is a polynomial in frequencies and momenta of degree
$\delta_{\Gamma}$, with the convention that $\omega \sim k^{2}$.

For the model (\ref{action}), dimensional analysis should be augmented by
the following additional considerations \cite{ANU}:

(i) All the 1-irreducible Green functions without the response fields
($N_{v'}=N_{\phi'}=0$) involve closed circuits of retarded propagators,
vanish identically, and therefore require no counterterms \cite{Book3}.

(ii) If for some reason a number of external momenta occurs as an overall
factor in all the diagrams of a given Green function, the real index of
divergence $\delta_{\Gamma}'$ is smaller than $\delta_{\Gamma}$ by the
corresponding number of unities \cite{turbo,Book3}. In the model
(\ref{action}) the field $\phi$ enters the vertex $\phi'(v\partial)\phi$
only in the form of spatial derivative, which reduces the real index of
divergence:
\begin{equation}
\delta_{\Gamma}' = \delta_{\Gamma}- N_{\phi}.
\label{real}
\end{equation}
The field $\phi$ enters
the counterterms only in the form of the derivative $\partial\phi$.
In particular, for the 1-irreducible function
$\langle\phi'\phi\rangle_{\rm 1-ir}$
one obtains $\delta_{\Gamma} = 2$, $\delta_{\Gamma}' = 0$. Thus the
counterterm $\phi' \partial_{t}\phi$, allowed by dimensional analysis,
is in fact forbidden, and the only possible structure is
$\phi' \partial^{2}\phi$.

(iii) Galilean invariance of the model (\ref{action}) requires that the
contributions of the counterterms be also invariant. In particular,
this means that the covariant derivative (\ref{nabla}) enters the
counterterms as a whole. As a consequence, the counterterm required
for the 1-irreducible function $\langle \phi' v \phi\rangle_{\rm 1-ir}$ with
$\delta_{\Gamma} = 1$, $\delta_{\Gamma}' = 0$, necessarily has the form
$\phi'(v\partial)\phi$ and appears in the combination
$\phi'\nabla_{t}\phi$ with the counterterm $\phi' \partial_{t}\phi$
discussed above. Hence, it is also forbidden.

Similarly, the divergences in the functions $\langle v'v\rangle_{\rm 1-ir}$
with $\delta_{\Gamma} = 2$ and $\langle v'vv\rangle_{\rm 1-ir}$ with
$\delta_{\Gamma}=1$
can be eliminated by the two counterterms: $v'\partial^{2} v$ and the
combination $v'\nabla_{t}v$. In fact, the latter is also forbidden by the
{\it generalized} Galilean invariance with the time-dependent
transformation velocity parameter ${\bf w}(t)$ \cite{DeDom,GGp}:
\begin{eqnarray}
{\bf v}_{w}(x) &=& {\bf v}(x_{w}) - {\bf w}(t), \quad
\Phi_{w}(x) = \Phi(x_{w}),
\nonumber \\
x &=& \{t,{\bf x}\}, \quad x_{w} =\{t,{\bf x}+{\bf u}(t)\},
\nonumber \\
{\bf u}(t) &=& \int^{t} {\bf w}(t') dt'.
\label{GGi}
\end{eqnarray}
Here $\Phi$ denotes the three fields $v',\phi',\phi$. The action functional
is {\it not} invariant  with respect to such a transformation:
${\cal S}(\Phi_{w})={\cal S}(\Phi) + v'\partial_{t} w$. One can show,
however, that the generating functional
of the 1-irreducible Green functions transforms in the identical way,
${\Gamma}(\Phi_{w})={\Gamma}(\Phi) + v'\partial_{t} w$. Since in general
$ {\Gamma}(\Phi) = {\cal S}(\Phi)$ plus the diagrams with loops (which
contain all the UV divergences), the counterterms appear invariant under
(\ref{GGi}). This excludes the counterterm $v'\nabla_{t}v$, invariant
with respect to conventional Galilean transformation with a constant
${\bf w}$, but not invariant with respect to (\ref{GGi}). More detailed
discussion of the uses of the generalized Galilean transformation,
especially for composite fields, can be found in \cite{turbo,Book3,GGp}.

(iv) Expressions (\ref{lines}) show that the propagators
$\langle v'\phi \rangle_{0}$ and $\langle v \phi \rangle_{0}$ contain the
factor $c_{0}^{2}$, while $\langle v'\phi \rangle_{0}$ contains $c_{0}^{4}$.
These factors appear as external numerical factors in any diagram
involving these propagators, and its real index of divergence reduces by
the corresponding number of unities. In particular, any diagram of the
1-irreducible function with $N_{\phi'}>N_{\phi}$ contains the factor
$c_{0}^{2(N_{\phi'}-N_{\phi})}$. It then follows that the counterterm to
the 1-irreducible function $\langle \phi'v \rangle_{\rm 1-ir}$ with
$\delta_{\Gamma}=3$
necessarily reduces to $c_{0}^{2}\phi' (\partial v)$, while the structures
$\phi' \partial^{2} (\partial v)$ {\it etc.} are forbidden. Another
consequence is finiteness of the
function $\langle \phi'vv \rangle_{\rm 1-ir}$ with
$\delta_{\Gamma}=2$. Each diagram
of this function contains the factor $c_{0}^{2}$, which forbids the
counterterms like $\phi' (\partial v)(\partial v)$ {\it etc.}, while the
remaining structure $c_{0}^{2}  \phi' v^{2}$ is forbidden by the Galilean
symmetry.

Using all these considerations one can check that all the UV divergences in
the model (\ref{action}) are removed by the counterterms of the form
\begin{eqnarray}
v_{i}'\partial^{2} v_{i}, \ v_{i}' \partial_{i} \partial_{k} v_{k}, \
v_{i}'\partial_{i}\phi, \ c_{0}^{2} \phi'\partial_{i}v_{i}, \
\phi'\partial^{2} \phi.
\label{counter}
\end{eqnarray}
All these structures are present in the extended action functional
(\ref{action}) with $v_{0}>0$, so the model is multiplicatively
renormalizable.

Like for the incompressible case \cite{Two}, for $d=2$ a new UV divergence
arises in the function $\langle v'v'\rangle_{\rm 1-ir}$, and a new
counterterm
$v'\partial^{2} v'$ should be included. This case requires special treatment,
and in the following we assume $d>2$. Then the renormalized action functional
has the form
\begin{widetext}
\begin{eqnarray}
{\cal S}^{R}(\Phi) &=& \frac{1}{2} v_{i}'D^{f}_{ik}  v_{k}' +
v_{i}' \left\{ -\nabla_{t} v_{i} +
Z_{1} \nu [\delta_{ik}\partial^{2}-\partial_{i}\partial_{k}] v_{k} +
Z_{2} u \nu\partial_{i}\partial_{k} v_{k} - Z_{4}\partial_{i} \phi \right\} +
\nonumber \\
&+& \phi' \left[ -\nabla_{t}\phi  + Z_{3}v \nu \partial^{2} \phi -
Z_{5} c^{2}(\partial_{i}v_{i}) \right].
\label{Raction}
\end{eqnarray}
\end{widetext}

Here $g,\nu,u,v,c$ are renormalized counterparts of the original (bare)
parameters (with the subscript ``o''), the function $D^{f}$ is expressed
in renormalized parameters using the relation $g_{0}\nu_0^{3} =
g\mu^y\nu^3$, the reference scale (or the ``normalization mass'') $\mu$
is an additional free parameter of the renormalized theory; the
renormalization constants $Z_{i}$ depend only on the completely
dimensionless parameters $g,u,v,\alpha,d,y$. The renormalized action
(\ref{Raction}) is obtained from the original one (\ref{action})
by the renormalization of the fields $\phi\to Z_{\phi}\phi$,
$\phi'\to Z_{\phi'}\phi'$ and the parameters
\begin{eqnarray}
g_{0} &=& g \mu^y Z_{g}, \quad \nu_0=\nu Z_{\nu}, \quad
u_{0}= u Z_{u},
\nonumber \\
v_{0} &=& v Z_{v}, \quad c_{0}= c Z_{c}.
\label{mult}
\end{eqnarray}
The renormalization constants in (\ref{Raction}) and (\ref{mult}) are
related as
\begin{eqnarray}
Z_{\nu} &=& Z_{1}, \quad Z_{u}=Z_{2}Z_{1}^{-1},
\nonumber \\
Z_{v} &=&  Z_{3}Z_{1}^{-1}, \quad Z_{\phi}=Z_{\phi'}^{-1}=Z_{4},
\nonumber \\
 Z_{c} &=& (Z_{4}Z_{5})^{1/2}, \quad Z_{g}= Z_{\nu}^{-3}.
\label{relat}
\end{eqnarray}
The last relation follows from the absence of renormalization of the
non-local term of the random force in (\ref{Raction}); for the same
reason the parameters $m,\alpha$ from the correlation function  (\ref{force}) are
not renormalized: $Z_{m}=Z_{\alpha}=1$. No renormalization of the fields
$v,v'$ is needed: $Z_{v}=Z_{v'}=1$ due to the absence of renormalization
of the term $v'\nabla_{t}v$.

The renormalization constants are found from the requirement that the
Green functions of the renormalized model (\ref{Raction}), when expressed
in renormalized variables, be UV finite (in our case, be finite at $y\to0$).
In the minimal subtraction (MS) scheme, which is always used in what follows,
they have the form ``$Z=1+$ only poles in $y$.''  The calculation in
the first order in $g$ (one-loop approximation) gives \cite{ANU}
\begin{widetext}
\begin{eqnarray}
Z_{1} &=& 1+ \frac{\hat g}{y}\, A, \quad Z_{2}= 1+ \frac{\hat g}{uy}\, B,
\quad 
Z_{3} = 1+ \frac{\hat g}{y}\, \frac{(d-1)}{2dv(v+1)} -
\frac{\alpha\hat g}{y}\, \frac{(u-v)}{2duv(u+v)^{2}}, \quad
\nonumber \\
Z_{4}
&=& 1+ \frac{\hat g}{y}\, \frac{(d-1)}{2d(u+1)(v+1)}, \quad
Z_{5}= 1,
\label{Z}
\end{eqnarray}
\end{widetext}
with corrections of order ${\hat g}^{2}$ and higher.
Here we passed to the new coupling constant
\begin{eqnarray}
{\hat g}=g S_{d}/(2\pi)^{d},
\label{ghat}
\end{eqnarray}
where
\begin{eqnarray}
S_d= 2\pi^{d/2}/\Gamma (d/2)
\label{Sd}
\end{eqnarray}
is the surface area of the unit sphere in $d$-dimensional space
and $\Gamma(\cdots)$ is Euler's Gamma function, and denoted
\begin{widetext}
\begin{eqnarray}
A = \frac{d(d-1)u^{2} - 2(d^{2}+d-4)u-d(d+3)}{4d(d+2)(1+u)^{2}} +
\frac{\alpha(1-u)}{2du(1+u)^{2}},
\quad
B = (1-d)\, \frac{(d-1)u^{2}+(d+4)u+1}{2d(d+2)(1+u)^{2}}.
\label{AB}
\end{eqnarray}
\end{widetext}

One important technical remark follows. The renormalization constants
in the MS scheme do not depend on the dimensional parameter $c_{0}$.
On the other hand, all the propagators (\ref{lines}), and hence all
the Feynman diagrams, have a well-defined limit for $c_{0}\to 0$.
Thus in the calculation of the constants $Z_{1}$--$Z_{4}$ one can simply
set $c_{0}=0$ in (\ref{lines}) and (\ref{energies}). Then the propagators
$\langle \phi v' \rangle_{0}$, $\langle v\phi \rangle_{0}$,
$\langle \phi\phi \rangle_{0}$ vanish, while the form of the others
drastically simplifies. In the calculation of the constant $Z_{5}$ in front
of the term $c_{0}^{2} \phi'(\partial v)$ it is sufficient to take into
account the diagrams with one and only one propagator
$\langle \phi v' \rangle_{0}$ or $\langle v\phi \rangle_{0}$. Then the
needed $c_{0}^{2}$ appears as an external factor, and in the remaining
expression one can set $c_{0}=0$.

To avoid possible misunderstanding, we stress that we are interested in the
model with finite and arbitrary $c_{0}$, and that more involved
calculation with the full-scale propagators (\ref{lines}) would give the
same results (\ref{Z}), (\ref{AB}) for the renormalization constants.
In this respect, the parameter $c_{0}$ is similar to
$\tau \propto T-T_{c}$, deviation of the temperature from its critical
value, in models of critical behavior: in the MS scheme, the
renormalization constants do not depend on it and can be calculated
directly at the critical point $\tau=0$.

The simple expression $Z_{5}=1$ results from the cancellation of nontrivial
contributions from three Feynman diagrams; we see no reason to expect that
it is valid to all orders in $g$.

\subsection{RG equations and RG functions} \label{sec:NScomp4}

Let us recall a simple derivation of the RG equations; detailed
discussion can be found in  \cite{Zinn,Book3}.
The RG equations are written for the renormalized
correlation functions $G^{R} =\langle \Phi\cdots\Phi\rangle_{R}$, which
differ from the original (unrenormalized) ones
$G =\langle \Phi\cdots\Phi\rangle$ only by normalization and choice of the
parameters, and thus can be equally used for the analysis of the critical
behavior. The relation
${\cal S}_{R} (\Phi,e,\mu) = {\cal S} (Z_{\Phi}\Phi,e_{0})$ between the
functionals (\ref{action}) and (\ref{Raction}) results in the relations
\begin{equation}
G(e_{0},\dots) = Z_{\phi}^{N_{\phi}} Z_{\phi'}^{N_{\phi'}}
G^{R}(e,\mu,\dots)
\label{multi}
\end{equation}
between the correlation functions. Here, as usual,
$N_{\phi}$ and $N_{\phi'}$ are the numbers of corresponding fields
entering into $G$ (we recall that in our model $Z_{v}=Z_{v'}=1$);
$e_{0}=\{\nu_0, g_{0}, u_{0}, v_{0} \}$  is the full set of
bare parameters and $e=\{ \nu, g, u, v  \}$ are their renormalized
counterparts; the ellipsis stands for the other arguments
(times, coordinates, momenta {\it etc.}).

We use $\widetilde{\cal D}_{\mu}$ to denote the differential operation
$\mu\partial_{\mu}$ for fixed $e_{0}$ and operate on both sides of the
equation (\ref{multi}) with it. This gives the basic RG differential
equation:
\begin{equation}
\left\{ {\cal D}_{RG} + N_{\phi}\gamma_{\phi} +
N_{\phi'}\gamma_{\phi'} \right\} \,G^{R}(e,\mu,\dots) = 0,
\label{RG1}
\end{equation}
where ${\cal D}_{RG}$ is the operation $\widetilde{\cal D}_{\mu}$
expressed in the renormalized variables:
\begin{equation}
{\cal D}_{RG}= {\cal D}_{\mu} + \beta_{g}\partial_{g} +
\beta_{u}\partial_{u} + \beta_{v}\partial_{v}
- \gamma_{\nu}{\cal D}_{\nu}- \gamma_{c}{\cal D}_{c} .
\label{RG2}
\end{equation}
Here we have written ${\cal D}_{x} \equiv x\partial_{x}$ for any variable
$x$. The anomalous dimension $\gamma_{F}$ of a certain quantity $F$
(a field or a parameter) is defined as
\begin{equation}
\gamma_{F}= Z_{F}^{-1} \widetilde{\cal D}_{\mu} Z_{F} =
\widetilde{\cal D}_{\mu} \ln Z_F ,
\label{RGF1}
\end{equation}
and the $\beta$ functions for the three dimensionless coupling
constants $g$, $u$ and $v$ are
\begin{eqnarray}
\beta_{g} &=& \widetilde {\cal D}_{\mu} g = g\,[-y-\gamma_{g}],
\nonumber \\
\beta_{u} &=& \widetilde {\cal D}_{\mu} u = -u\gamma_{u},
\nonumber \\
\beta_{v} &=& \widetilde {\cal D}_{\mu} v = -v\gamma_{v},
\label{betagw}
\end{eqnarray}
where the second equalities result from the definitions and the
relations (\ref{multi}).

From the relations (\ref{relat}) we obtain
\begin{eqnarray}
\beta_{g} &=& g\, [-y+3\gamma_{1}],
\nonumber \\
\beta_{u} &=& u\, [\gamma_{1}-\gamma_{2}],
\nonumber \\
\beta_{v} &=& v\, [\gamma_{2}-\gamma_{3}],
\label{gammas2}
\end{eqnarray}
and for the anomalous dimensions we have
\begin{eqnarray}
\gamma_{\phi} &=& - \gamma_{\phi'} = \gamma_{4},
\quad
\gamma_{c} =(\gamma_{4}+\gamma_{5})/2,
\quad
\gamma_{\nu}= \gamma_{1},
\nonumber \\
\gamma_{v} &=& \gamma_{v'}=\gamma_{\alpha}=\gamma_{m}=0.
\label{gammas3}
\end{eqnarray}
The relations in the second line follow from the absence of renormalization
of the corresponding fields and parameters; see the remarks below equation
(\ref{relat}).

In the MS scheme all the renormalization constants have the form
\begin{eqnarray}
Z_{F} = 1 + \sum_{n=1}^{\infty} z^{(n)} y^{-n},
\label{ZMS}
\end{eqnarray}
where the coefficients $z^{(n)}$ do not depend on $y$. Then from
the definition and the expressions (\ref{betagw}) it follows that
the corresponding anomalous dimension is determined solely by the
first-order coefficient:
\begin{eqnarray}
\gamma_{F}=-{\cal D}_{g} z^{(1)},
\label{E}
\end{eqnarray}
see {\it e.g.} the discussion \cite{Zinn,Book3}. Then in the one-loop
approximation from the explicit expressions (\ref{Z}) one finds:
\begin{eqnarray}
\gamma_{1} &=& - A \hat{g}, \quad \gamma_{2}= - B \hat{g} /u,
\nonumber \\
\gamma_{3} &=& \hat{g}\, \frac{(d-1)}{2dv(v+1)} +
{\alpha\hat g}\, \frac{(u-v)}{2duv(u+v)^{2}},
\nonumber \\
\gamma_{4} &=& \hat{g}\, \frac{(1-d)}{2d(u+1)(v+1)}, \quad
\gamma_{5}= 0
\label{gammasE}
\end{eqnarray}
with $A$ and $B$ from (\ref{AB}) and
the corrections of order $\hat{g}^{2}$ and higher.

\subsection{The IR attractive fixed point} \label{sec:NScomp5}

It is well known that possible IR asymptotic regimes of a renormalizable
field theoretic model are associated with IR attractive fixed points of the
corresponding RG equations. The coordinates $g_{*}$
of the fixed points are found from the equations
\begin{equation}
\beta_{i} (g_{*}) =0,
\label{points}
\end{equation}
where $g=\{g_i\}$ is the full set of coupling constants and $\beta_{i}$
are the corresponding $\beta$ functions.
The type of a fixed point is determined by the matrix
\begin{equation}
\Omega_{ij}=\partial\beta_{i}/\partial g_{j}|_{g=g_{*}}.
\label{Omega}
\end{equation}
For the IR stable fixed points the matrix $\Omega$ is positive, {\it i.e.},
the real parts of all its eigenvalues are positive.

In our model, $g = \{\hat g, u, w\}$, and the $\beta$ functions are given
be the relations (\ref{betagw}) and the explicit one-loop expressions
(\ref{gammasE}). We do not include the dimensionless parameter $\alpha$
into the list of coupling constants, because it is not renormalized
($\alpha_{0}=\alpha$ and $Z_{\alpha}=1$) and the corresponding function
$\beta_{\alpha}=-\alpha\gamma_{\alpha}$ vanishes identically. Thus the
equation $\beta_{\alpha}=0$ imposes no restriction on the value of $\alpha$,
and it remains a free parameter.

Analysis of the expressions (\ref{betagw}), (\ref{gammasE}) and (\ref{AB})
shows that in the physical range of parameters ($\hat g, u, v, \alpha>0$)
there is only one IR attractive fixed point with the coordinates
\begin{equation}
\hat g_{*} = \frac{4dy}{3(d-1)}, \quad u_{*}=v_{*}=1,
\label{FP}
\end{equation}
with possible higher-order corrections in $y$.

Let us briefly explain derivation of (\ref{FP}). Any fixed point with
$\hat g_{*} =0$ cannot be IR attractive, because one of the eigenvalues
of the matrix $\Omega$ coincides with the diagonal element
$\partial_{g} \beta_{g} = -y <0$. For $\hat g_{*} \ne 0$ from the
equation $\beta_{g} =0$ we immediately find the relation
$\gamma_{1}^{*}= \gamma_{\nu}^{*}= y/3$, valid to all orders in $y$
(here and below $\gamma_{F}^{*}= \gamma_{F}(g_{*})$ for any $F$
is the value of the anomalous dimension at the fixed point in question).
Substituting this relation into the equation $\beta_{u} =0$ gives the
equation for $u_{*}$ with the only positive solution $u_{*}=1$.
Substituting it into the equation $\beta_{g} =0$ gives the value of
$\hat g_{*}$ (it is important here that the functions $\beta_{g}$ and
$\beta_{u}$ in the one-loop approximation do not depend on $v$).
Finally, substituting the known values of $\hat g_{*}$ and $u_{*}$ into
the relation $\beta_{v}=0$ gives the equation for $v_{*}$ with the only
positive solution $v_{*}=1$. Now it is easy to see that the matrix
(\ref{Omega}) at the fixed point (\ref{FP}) is triangular, so that its
eigenvalues coincide with the diagonal elements and are easily calculated
from the explicit expressions (\ref{gammasE}). They are positive for all
$y>0$, $\alpha>0$ and $d>2$.

It is also worth noting that the so-called ``RG  flows'' (solutions to the
RG equations for the RG-invariant or ``running'' coupling constants) cannot
leave the physical range $\hat g, u, v >0$ (for the physical initial data).
This follows from the fact that all the $\beta$ functions vanish for $g=0$
and that the functions $\beta_{u}$ and $\beta_{v}$ are negative for $u=0$
and $v=0$, respectively:
\[ \beta_{u}|_{u=0} = - \hat g\, \frac{(d-1)}{2d(d+2)},
\quad
\beta_{v}|_{v=0} = - \hat g\, \left\{ \frac{(d-1)}{2d}
+ \frac{1}{du^{2}} \right\}. \]
It then follows that the IR asymptotic behavior of the Green functions in
our model can be governed only by the fixed point (\ref{FP}): even if some
other fixed points exist in the unphysical range, they cannot be reached by
the RG flow.

Changing to the new variable $f=1/u$ one can find another fixed point with
$f_{*}=0$ and $\hat g_{*}= 4(d+2)y/3(d-1)$. From the explicit form of the
propagators (\ref{lines}) it follows, that the limit $u\to\infty$
corresponds to the purely transverse velocity field, while the scalar field
decouples. The point is unstable (it is a saddle point) in agreement with
the analysis of Refs. \cite{Lviv,KompOp,NV} which shows that the
leading-order correction in the Mach number to the incompressible scaling
regime destroys its stability (in the RG terminology, it is relevant in
the sense of Wilson).

\subsection{IR behavior and the critical dimensions}
\label{sec:NScomp6}

It follows from the solution of the RG equation (\ref{RG1}) that when
an IR fixed point is present, the leading term of the IR asymptotic
behavior of the Green function $G^{R}$ satisfies the equation
(\ref{RG1}) with the replacement $g\to g_{*}$ for the full set of the
couplings; see {\it e.g.} the monograph \cite{Book3}. In our case this
gives
\begin{equation}
\left\{
{\cal D}_{\mu} - \gamma_{\nu}^{*}{\cal D}_{\nu}- \gamma_{c}^{*}{\cal D}_{c}
+ \sum_{\Phi} N_{\Phi}\gamma_{\Phi}^{*}  \right\} \, G^{R} = 0.
\label{RGF}
\end{equation}
We recall that ${\cal D}_{x}\equiv x\partial_{x}$ for any variable $x$,
$\gamma_{F}^{*}$ is the fixed-point value of the anomalous dimension
$\gamma_{F}$, and the summation over all types of the fields $\Phi$ is
implied. In the one-loop approximation, from (\ref{gammasE}) and (\ref{FP})
we obtain
\begin{eqnarray}
\gamma_{\nu}^{*} &=& y/3 \ {\rm (exact)}, \quad
\gamma_{\phi}^{*}=-\gamma_{\phi'}^{*} =-y/6+O(y^{2}),
\nonumber \\
\gamma_{c}^{*} &=& -y/12 +O(y^{2}).
\label{Anom}
\end{eqnarray}

Canonical scale invariance is expressed by the two equations
\begin{eqnarray}
\left\{ \sum_{F} d^{k}_{F} {\cal D}_{F} - d^{k}_{G} \right\} \, G^{R} = 0,
\nonumber \\
\left\{ \sum_{F} d^{\omega}_{F} {\cal D}_{F} -
d^{\omega}_{G} \right\} \, G^{R} = 0,
\label{Ck}
\end{eqnarray}
with the summations over all the arguments of the function $G^{R}$. From
table~1 we obtain
\begin{eqnarray}
\left\{ - {\cal D}_{\bf x} +{\cal D}_{\mu} +{\cal D}_{m} -2 {\cal D}_{\nu}
- {\cal D}_{c} - \sum_{\Phi} N_{\Phi} d^{k} _{\Phi}
\right\} \, G^{R} = 0,
\nonumber \\
\left\{ - {\cal D}_{t} + {\cal D}_{\nu} + {\cal D}_{c}
- \sum_{\Phi} N_{\Phi} d^{\omega} _{\Phi} \right\} \, G^{R} = 0,
\nonumber \\ {}
\label{Ck2}
\end{eqnarray}
where the dimensions $d^{k,\omega}_{\Phi}$ of the fields are also given in
the table. Each of the equations (\ref{RGF}), (\ref{Ck2}) describes the
scaling with dilatation of the variables whose derivatives enter the
differential operator. One is interested in the scaling with fixed
``IR irrelevant'' parameters $\mu$ and $\nu$; see \cite{turbo,UFN,Book3}.
In order to derive the corresponding scaling equation one has to combine
(\ref{RGF}), (\ref{Ck2}) such that the derivatives with respect to these
parameters be eliminated; this gives:
\begin{eqnarray}
\left\{ - {\cal D}_{\bf x} + \Delta_{t} {\cal D}_{t}
+ \Delta_{c} {\cal D}_{c}+ \Delta_{m} {\cal D}_{m} -
\sum_{\Phi} N_{\Phi} \Delta_{\Phi} \right\} \, G^{R} = 0
\nonumber \\ {}
\label{KS}
\end{eqnarray}
with
\begin{equation}
\Delta_{F} = d^{k}_{F}+ \Delta_{\omega}d^{\omega}_{F} + \gamma_{F}^{*},
\quad \Delta_{\omega} = - \Delta_{t} =  2-\gamma_{\nu}^{*}.
\label{Krit}
\end{equation}
Here $\Delta_{F}$ is the critical dimension of the quantity $F$ (following
\cite{turbo,UFN,Book3} we use this term to distinguish it from canonical
dimensions), while $\Delta_{t}$ and $\Delta_{\omega}$ are the critical
dimensions of the time and the frequency.

From table~1 and expressions (\ref{Anom}) we obtain
\begin{equation}
\Delta_{v}=1-y/3, \quad \Delta_{v'}= d- \Delta_{v}, \quad
\Delta_{\omega}=2-y/3, \quad \Delta_{m}=1
\label{KritEx}
\end{equation}
(these results are exact due to $\gamma_{\nu}^{*}=y/3$ and
$\gamma^{*}_{v,v',m}=0$) and
\begin{equation}
\Delta_{\phi}=d-\Delta_{\phi'}=2-5y/6+O(y^{2}), \quad
\Delta_{c}= 1- 5y/12 +O(y^{2}).
\label{Krit2}
\end{equation}
We note that the analogs of the expressions (\ref{KritEx}), (\ref{Krit2})
in Ref. \cite{ANU} contain a few misprints.

Surprisingly enough, all the results (\ref{FP}), (\ref{Anom}),
(\ref{KritEx}), (\ref{Krit2}) are independent on $\alpha$ (and some of them
do not depend on $d$). They are valid for all $\alpha>0$, but the case
$\alpha\to\infty$ (purely potential random force) requires special attention.
To study this limit, one should pass to the new couplings $g'=g\alpha$,
$b=1/\alpha$ and then set $b=0$ at finite $g'$. This gives
\begin{equation}
\beta_{g'}=-yg' , \quad \beta_{u}=g'\frac{(u-1)}{2du(1+u)^{2}}, \quad
\beta_{v}=g' \frac{(v-u)}{du(u+v)^{2}}.
\label{potent}
\end{equation}
The system (\ref{potent}) has no IR attractive fixed point, because from
$\beta_{g'}=0$ it necessarily follows that $g'=0$, and such a point cannot
be IR attractive due to $\partial_{g'} \beta_{g'}=-y<0$. In principle,
the needed fixed point with $g'_{*} \sim y^{1/2}$ can appear on the two-loop
level, if the term of order $(g')^{3}$ appears in $\beta_{g'}$. Then the
results (\ref{KritEx}) remain valid, while (\ref{Krit2}) should be revised.


\section{Passive scalar fields: Renormalization, RG functions and fixed
point} \label{sec:PD}

\subsection{The models and their field theoretic formulation}
\label{sec:PD1}

There are two main types of diffusion-advection problems for the
compressible velocity field \cite{LL}.
Passive advection of a density field $\theta(x)\equiv \theta(t,{\bf x})$
(say, the density of a pollutant) is described by the equation
\begin{equation}
\partial _t\theta+ \partial_{i}(v_{i}\theta)=\kappa _0 \partial^{2} \theta+f,
\label{density1}
\end{equation}
while the advection of a ``tracer'' (temperature, specific entropy, or
{\it concentration} of the impurity particles) is described by
\begin{equation}
\partial _t\theta+ (v_{i}\partial_{i})\theta=\kappa _0\partial^{2} \theta+f.
\label{tracer1}
\end{equation}
Here $\partial _t \equiv \partial /\partial t$,
$\partial _i \equiv \partial /\partial x_{i}$, $\kappa_0$ is the molecular
diffusivity coefficient, $\partial^{2}=\partial _i\partial _i$
is the Laplace operator, ${\bf v}(x)$ is the velocity field,
and $f\equiv f(x)$ is a Gaussian noise with zero mean and given covariance,
\begin{equation}
\langle f(x)f(x') \rangle = \delta(t-t')\, C({\bf r}/L), \quad
{\bf r}= {\bf x} - {\bf x}'.
\label{noise}
\end{equation}
Here $C({\bf r}/L)$ is some function finite at $({\bf r}/L)\to 0$ and rapidly
decaying for $({\bf r}/L)\to\infty$. In the following, we do not distinguish
the integral scale $L$, related to the noise, and its analog $L=m^{-1}$ in
the correlation function  of the stirring force (\ref{power}). Without
loss of
generality, one can set $C(0)=1$ (the coefficient can be absorbed by
rescaling of $\theta$ and $f$). The noise mimics the effects of initial
and/or boundary conditions: it maintains the steady state of the system
and serves as the source of the large-scale anisotropy. (The latter term
means that the anisotropy is introduced at scales of order $L$, while the
statistics of the velocity field remains isotropic. The case of anisotropic
velocity statistics is discussed, within the RG+OPE approach, in Refs.
\cite{Aniso}.) In more realistic formulations, the noise can arise
from an imposed linear gradient of the (temperature) field. It
turns out, however, that the specific form of the random stirring is
unimportant, and in the following we use the artificial noise with the
correlation function  (\ref{noise}).

In the absence of the noise, equation (\ref{density1}) has the form of a
continuity equation (conservation law); $\theta$ being the density of a
corresponding conserved quantity. For (\ref{tracer1}), the conserved
quantity is the auxiliary (response) field $\theta'$, which appears in the
field-theoretic formulation of the problem; see below. If the function in
(\ref{noise}) is chosen such that its Fourier transform $C({\bf k})$
vanishes at ${\bf k}=0$, the fields $\theta$ or $\theta'$
remain to be conserved in the statistical sense in the presence of the
external stirring.

The models (\ref{density1}) and (\ref{tracer1}) were thoroughly studied
for the case of Kraichnan's rapid-change model \cite{VM}--\cite{AG};
the case of Gaussian velocity statistics with finite correlation time
was studied in \cite{AKens,AK2}.

The stochastic problem (\ref{density1}), (\ref{noise}) is equivalent
to the field theoretic model of the full set of fields
$\Phi\equiv\{\theta', \theta, v', v, \phi', \phi  \}$
with the action functional
\begin{equation}
{\cal S}_{\Phi}(\Phi)= {\cal S}_{\theta}(\theta', \theta, v) +
{\cal S}(v', v, \phi', \phi),
\label{Fact}
\end{equation}
where
\begin{eqnarray}
{\cal S}_{\theta} (\theta', \theta, v) = \frac{1}{2} \theta' D_{f} \theta' +
\theta' \left\{ - \partial_{t}\theta -
\partial_{i}(v_{i}\theta) +\kappa _0 \partial^{2} \theta \right\}
\nonumber \\
\label{Dact}
\end{eqnarray}
is the De Dominicis--Janssen action for the stochastic problem
(\ref{density1}),
(\ref{noise}) at fixed ${\bf v}$, while the second term is given by
(\ref{action}) and the represents the velocity statistics; $D_{f}$ is the
correlation function  (\ref{noise}), and, as usual, all the required
integrations and summations over the vector indices are implied.

In addition to (\ref{lines}), the diagrammatic technique in the full
problem involves two propagators
\begin{eqnarray}
\langle \theta \theta' \rangle _0=  \langle \theta' \theta \rangle_0^* &=&
\frac{1} {-{\rm i}\omega +\kappa _0 k^2},
\nonumber \\
\langle \theta \theta \rangle _0 &=& \frac {C({\bf k})}
{\omega^{2} +\kappa_0^{2} k^4},
\label{lines3}
\end{eqnarray}
and the new vertex $-\theta'\partial_{i}(v_{i}\theta)=
V_{i} \theta' v_{i}\theta$. In the momentum representation,
the vertex factor $V_{i}$ in the diagrams has the form
\begin{eqnarray}
V_{i} ({\bf k}) = {\rm i}  k_{i},
\label{vertex1}
\end{eqnarray}
where ${\bf k}$ is the momentum argument of the field $\theta'$ (using
integration by parts, the derivative at the vertex can be moved onto the
field $\theta'$).

The problem (\ref{tracer1}), (\ref{noise})  corresponds to the action
(\ref{Fact}), where the part ${\cal S}_{\theta}$ is given by
\begin{eqnarray}
{\cal S}_{\theta} (\theta', \theta, v) = \frac{1}{2} \theta' D_{f} \theta' +
\theta' \left\{ - \partial_{t}\theta -
(v_{i}\partial_{i})\theta +\kappa _0 \partial^{2} \theta \right\}.
\nonumber \\
\label{Tact}
\end{eqnarray}
The propagators are given by the same expressions (\ref{lines3}),
while the vertex factor (\ref{vertex1}) is replaced with
\begin{eqnarray}
V_{i} ({\bf k}) = - {\rm i}  k_{i},
\label{vertex2}
\end{eqnarray}
where ${\bf k}$ is the momentum argument of the field $\theta$.

\subsection{UV renormalization and all that} \label{sec:PD2}

Canonical dimensions of the new fields and parameters that appear in the
models  (\ref{Fact}), (\ref{Dact}), (\ref{Tact}) are given in table~1,
where we introduced a new dimensionless parameter
$w_{0}=\kappa_{0}/\nu_{0}$ with $\nu_{0}$ from (\ref{NS}).

Now in the expression (\ref{index}) for the formal index of UV divergence
the summation runs over the full set of fields
$\Phi\equiv\{\theta', \theta, v', v, \phi', \phi  \}$. The rules (i)-(iv)
from section~\ref{sec:NScomp3} should be extended and augmented as follows:

(i) All the 1-irreducible Green functions without the response fields
$v',\phi',\theta'$ vanish identically and require no counterterms.

(ii) In the model (\ref{Tact}), the field $\theta$ enters the vertex
$-\theta'(v_{i}\partial_{i})\theta$ only in the form of derivative.
Then the expression (\ref{real}) for the real index of divergence
should be modified as
\begin{equation}
\delta_{\Gamma}' = \delta_{\Gamma}- N_{\phi} - N_{\theta}.
\label{realT}
\end{equation}
In the model (\ref{Dact}), the derivative at the vertex
$-\theta'\partial_{i}(v_{i}\theta)$ can be moved onto the field $\theta'$
using integration by parts, and the real index becomes
\begin{equation}
\delta_{\Gamma}' = \delta_{\Gamma}- N_{\phi} - N_{\theta'}.
\label{realD}
\end{equation}
Since the field $\theta$ in model (\ref{Tact}) and $\theta'$ in model
(\ref{Tact})
can enter the counterterms only in the form of spatial derivatives,
the counterterm $\theta'\partial_{t}\theta$ to the 1-irreducible
Green function $\langle\theta'\theta\rangle_{\rm 1-ir}$ with
$\delta_{\Gamma}=2$, $\delta_{\Gamma}'=1$ is forbidden for the both models.

(iii) Another consequence of (ii) is that the counterterms to the
1-irreducible function $\langle\theta'v\theta\rangle_{\rm 1-ir}$ with
$\delta_{\Gamma}=1$, $\delta_{\Gamma}'=0$ necessarily reduce to the
form $\theta'\partial_{i}(v_{i}\theta)$ for the model (\ref{Dact})
and $\theta'(v_{i}\partial_{i})\theta$ for the model (\ref{Tact}).
Galilean symmetry requires, however, that these monomials enter the
counterterms in the form of invariant combinations
$\theta'[\partial_{t}\theta+\partial_{i}(v_{i}\theta)]$ and
$\theta'\nabla_{t}\theta$. Hence, they are also forbidden.

(iv) From the straightforward analysis of the Feynman diagrams it follows
that, for any 1-irreducible function, $N_{\theta'}- N_{\theta}=2N_{0}$,
where $N_{0}$ is the total number of bare propagators
$\langle\theta\theta\rangle_0$ entering the diagram. Clearly, no diagram
with $N_{0}<0$
can be constructed, so that the difference $N_{\theta'}- N_{\theta}$ is
an even non-negative integer for any nontrivial Green function. This fact,
a consequence of the linearity of the original stochastic equations
(\ref{density1}), (\ref{tracer1}) in the field $\theta$, appears crucial for
the renormalizability of the models (\ref{Dact}) and (\ref{Tact}). Indeed,
the total canonical dimension $d_{\theta}=-1$ is negative (in contrast to
most conventional field theoretic models), so that the index (\ref{realD})
increases with $N_{\theta}$, while (\ref{realT}) does not depend on
$N_{\theta}$. Without the restriction $N_{\theta}\le N_{\theta'}$, we
would face the infinity of superficially divergent functions
$\langle \theta'\theta\dots\theta\rangle_{\rm 1-ir}$, and hence the lack of
renormalizability.

Finally, we are left with the only superficially divergent 1-irreducible
Green function $\langle\theta'\theta\rangle_{\rm 1-ir}$ with the only
counterterm
$\theta'\partial^{2}\theta$. It is naturally reproduced as multiplicative
renormalization of the diffusion coefficient, $\kappa_{0}=\kappa Z_{\kappa}$.
No renormalization of the fields $\theta'$, $\theta$ is needed:
$ Z_{\theta'}=Z_{\theta}=1$. The renormalized analog of the action
functional (\ref{Dact}) has the form
\begin{equation}
{\cal S}_{\Phi}^{R}(\Phi)= {\cal S}_{\theta}^{R}(\theta', \theta, v) +
{\cal S}^{R}(v', v, \phi', \phi)
\label{FactR}
\end{equation}
with ${\cal S}^{R}$ from (\ref{Raction}) and
\begin{widetext}
\begin{equation}
{\cal S}^{R}_{\theta} (\theta', \theta, v) = \frac{1}{2} \theta' D_{f}
\theta' + \theta' \left\{ - \partial_{t}\theta -
\partial_{i}(v_{i}\theta) +\kappa Z_{\kappa} \partial^{2} \theta \right\},
\label{DactR}
\end{equation}
and similarly for (\ref{Tact}):
\begin{equation}
{\cal S}^{R}_{\theta} (\theta', \theta, v) = \frac{1}{2} \theta' D_{f}
\theta' + \theta' \left\{ - \partial_{t}\theta -
(v_{i}\partial_{i})\theta + \kappa Z_{\kappa} \partial^{2} \theta \right\}.
\label{TactR}
\end{equation}
\end{widetext}

It remains to note that, if the term with $D_{f}$ is omitted,
the models (\ref{Dact}) and (\ref{Tact})
can be mapped onto each other by means of the interchange
$\theta(t,{\bf x})  \leftrightarrow \theta'(t,{\bf x})$ and the reflection
$t\to-t$. In particular, this means that the renormalization constants
$Z_{\kappa}$ in (\ref{DactR}) and (\ref{TactR}) coincide to all orders of
the perturbation theory, because the correlator $D_{f}$ does not appear in
the relevant diagrams; see the next subsection.

\subsection{Explicit leading-order results. Fixed points and
scaling dimensions} \label{sec:PD3}

Let us turn to the explicit calculation of the renormalization constant
$Z_{\kappa}$ in the leading one-loop order; for definiteness, consider
the case of the density field (\ref{DactR}). The constant is found from the
requirement that the 1-irreducible Green function
$\langle\theta'\theta\rangle_{\rm 1-ir}$ be UV finite (that is, finite
at $y\to0$) when expressed in renormalized parameters. The corresponding
Dyson equation in the frequency--momentum representation reads:
\begin{equation}
\langle\theta'\theta\rangle_{\rm 1-ir}(\omega,p) = + {\rm i}\omega
- \kappa_0 p^{2} + \Sigma_{\theta'\theta} (\omega,p),
\label{Dyson}
\end{equation}
where the ``self-energy operator''  $\Sigma_{\theta'\theta}$ is given by
the infinite sum of 1-irreducible graphs. In the one-loop approximation it
has the form:
\begin{equation}
\Sigma_{\theta'\theta} =
\includegraphics [width=.13\textwidth,clip]{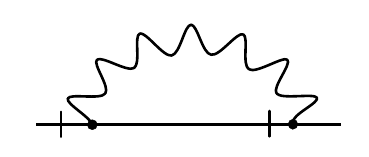}
\label{Self}
\end{equation}
where the wavy line denotes the bare propagator $\langle vv\rangle_0$ from
(\ref{lines}), the solid line with a slash
denotes the bare propagator $\langle \theta \theta' \rangle _0$ from
(\ref{lines3}), the slashed end corresponding to the field $\theta'$.
The dots with three attached fields $\theta'$, $\theta$, $v$ denote the
vertex (\ref{vertex1}).

In the leading-order approximation, the renormalization constant in the
bare term of (\ref{Dyson}) is taken only in the first order in $g$, that
is, $\kappa_{0} = \kappa Z_{\kappa} \simeq \kappa (1 + z^{(1)} g/y)$,
while in the diagram (\ref{Self}) all $Z$'s are replaced with unities.
Furthermore, we only need to know the divergent part of (\ref{Self}), which
is proportional to $p^{2}$ (see the preceding subsection). Thus we can set
$\omega=0$ in (\ref{Dyson}) and keep in the expansion in ${\bf p}$ of the
resulting integrand only the $p^{2}$ term. Like for the original NS model,
its divergent part is independent on $c_{0} \sim c$ and can be calculated
directly at $c=0$; see the discussion in subsec.~\ref{sec:NScomp3}. Then
the expression for (\ref{Self}) becomes:
\begin{equation}
 \Sigma_{\theta'\theta} = {\rm i} p_{s} \int\frac{d\omega'}{2\pi}
\int_{k>m}\frac{d{\bf k}}{(2\pi)^{d}}\, {\rm i} (p+k)_{l}
\frac{D_{sl} (\omega', {\bf k})}
{-{\rm i}\omega'+w\nu |{\bf p}+{\bf k}|^{2}} ,
\label{Self1}
\end{equation}
where
\begin{equation}
 D_{sl}(\omega', {\bf k}) = g\mu^y\nu^{3} \left\{
\frac {P^{\bot}_{sl} ({\bf k})}{ (\omega')^{2} + \nu^2 k^{4}}
+ \frac {\alpha P^{\parallel}_{sl} ({\bf k})}
{ (\omega')^{2} + u^{2}\nu^2 k^{4}} \right\}
\label{Self2}
\end{equation}
is the velocity correlation function  from (\ref{lines}) with the proper
substitutions, including $c=0$.

Integrations over the frequency are easily performed, for example,
\begin{widetext}
\begin{equation}
\int\frac{d\omega'}{2\pi}\,
\frac{1}{-{\rm i}\omega'+w\nu |{\bf p}+{\bf k}|^{2}}\,
\frac {1} { (\omega')^{2} + u^{2}\nu^2 k^{4}} =
 \frac {1} {2u\nu^{2}k^{2}(uk^{2}+w|{\bf p}+{\bf k}|^{2})}.
\label{Fre}
\end{equation}
\end{widetext}
In the terms containing the factor $p_{s}p_{l}$ one can immediately set
${\bf p}=0$ in (\ref{Fre}), while in the exceptional term with
$p_{s}k_{l}P^{\parallel}_{sl} ({\bf k})= p_{s}k_{s}$ one should expand
(\ref{Fre}) up to the linear term in ${\bf p}$:
\[ \frac {1} {uk^{2}+w|{\bf p}+{\bf k}|^{2}} =
\frac{1}{(u+w)k^{2}} \left\{ 1- \frac{2w}{(u+w)} \frac{(\bf pk)}{k^{2}}
\right\}. \]
With the aid of the formulas
\begin{widetext}
\begin{eqnarray}
\int\! {d{\bf k}} k_{i} f(k) =0, \
\int\! {d{\bf k}} \frac{k_{i}k_{s}}{k^{2}} f(k) =
\frac{\delta_{is}}{d}\, \int {d{\bf k}}\, f(k), \
\int\! {d{\bf k}} \frac{k_{i}k_{s}k_{l}k_{p}}{k^{4}} f(k) =
\frac{\delta_{is}\delta_{lp}+\delta_{il}\delta_{sp}+\delta_{ip}\delta_{sl}}
{d(d+2)} \int {d{\bf k}} f(k),
\label{tenz}
\end{eqnarray}
\end{widetext}
where $f(k)$ is any function depending only on $k=|{\bf k}|$, all the
resulting integrals are reduced to the scalar integral
\begin{eqnarray}
J(m)= \int_{k>m} {d{\bf k}}\, \frac{1}{k^{d+y}} = S_{d}
\frac{m^{-y}}{y}
\label{ska}
\end{eqnarray}
with $S_{d}$ from (\ref{Sd}).

Collecting all the terms gives
\begin{equation}
\Sigma_{\theta'\theta} = - \frac{{\hat g}} {2dy} \left(\frac{\mu}{m}\right)
^{y} \left\{ \frac{(d-1)}{(1+w)} + \frac{\alpha}{u(u+w)} -
\frac{2\alpha w} {u(u+w)^{2}}\right\}
\label{Selff}
\end{equation}
with ${\hat g}$ defined in (\ref{ghat}). Then the renormalization constant,
needed to cancel the pole in $y$ in (\ref{Dyson}), in the MS scheme should
be chosen as
\begin{equation}
Z_{\kappa} = 1 - \frac{{\hat g}} {2dwy}
\left\{ \frac{(d-1)}{(1+w)} + \frac{\alpha(u-w)}{u(u+w)^{2}}
\right\},
\label{Zk}
\end{equation}
while the corresponding anomalous dimension is
\begin{equation}
\gamma_{\kappa} =  \frac{\hat g} {2dw}
\left\{ \frac{(d-1)}{(1+w)} + \frac{\alpha(u-w)}{u(u+w)^{2}}
\right\},
\label{gk}
\end{equation}
with the corrections of the order ${\hat g}^{2}$ and higher.

The function $\beta_{w} = \widetilde {\cal D}_{\mu} w$ for the new
dimensionless parameter $w$ has the form
\begin{equation}
\beta_{w} = - w\gamma_{w} = w [\gamma_{\nu}-\gamma_{\kappa}],
\label{betaw}
\end{equation}
cf. equation (\ref{betagw}). Substituting the one-loop expressions
(\ref{FP}), (\ref{gk}) and the exact relation (\ref{Anom}) into the
equation $\beta_{w} =0$ gives, after some simple algebra, the equation
\begin{equation}
(w-1) [(d-1)(w+1)(w+2)+2\alpha] =0,
\label{kub}
\end{equation}
with the only positive solution $w_{*}=1$.

The corresponding new eigenvalue of the matrix (\ref{Omega}) coincides
with the diagonal element
\[\partial\beta_{w}/\partial w|_{g=g_{*}} = y [3(d-1)+\alpha]/ 6(d-1)>0,\]
because the functions (\ref{betagw}) do not depend on $w$.
We conclude that the fixed point with the coordinates
(\ref{FP}) and $w_{*}=1$ is IR attractive
in the full space of couplings $g,u,v,w$ and governs the IR asymptotic
behavior of the full-scale models (\ref{Dact}), (\ref{Tact}).

The critical dimensions of the fields $\theta$, $\theta'$ are obtained from
the data in table~1 and the expression (\ref{Krit}) for $\Delta_{\omega}$:
\begin{equation}
\Delta_{\theta}= -1+y/6, \quad \Delta_{\theta'}= d+1 -y/6.
\label{KriTet}
\end{equation}
These expressions are exact due to the absence of renormalization of
the fields $\theta$ and $\theta'$.

\section{Composite fields and their dimensions} \label{sec:Opera}

The key role in the following will be played by certain composite fields
(``composite operators'' in the quantum-field terminology). A local composite
operator is a monomial or polynomial constructed from the primary fields
$\Phi(x)$ and their finite-order derivatives at a single space-time point
$x=\{t,{\bf x}\}$. In the Green functions with such objects, new UV
divergences arise due to coincidence of the field arguments. They are
removed by additional renormalization procedure. As a rule, operators mix
in renormalization: renormalized operators are given by certain finite
linear combinations of the original monomials. However, in the following
only a simpler situation will be encountered, when the original operator
$F(x)$ and the renormalized one $F^{R}(x)$ are related by multiplicative
renormalization $F(x)= Z_{F} F^{R}(x)$ with the renormalization constant
of the form (\ref{ZMS}). Then the critical dimension of the operator is
given by the same expression (\ref{Krit}) and, in general, differs from
the simple sum of the dimensions of the fields and derivatives that enter
the operator.

The total canonical dimension of any 1-irreducible Green function $\Gamma$
with one operator $F(x)$ and arbitrary number of primary fields
(the formal index of UV divergence) is  given by
\begin{eqnarray}
\delta_{\Gamma} = d_{F} - \sum_{\Phi} N_{\Phi} d_{\Phi},
\label{indexo}
\end{eqnarray}
where $N_{\Phi}$ are the numbers of the fields entering into $\Gamma$,
$d_{\Phi}$ are their total canonical dimensions, $d_{F}$ is the canonical
dimension of the operator, and the summation over all types of the fields
is implied. Superficial UV divergences can be present only in the functions
$\Gamma$ with a non-negative integer $\delta_{\Gamma}$.

\subsection{Renormalization of the composite fields $\theta^{n}$.
Explicit leading-order results} \label{sec:CF}

Let us begin with the simplest case of the operators $F(x)=\theta^{n}(x)$
in the density model. Then $d_{F}=-n$ in (\ref{indexo}). Due to the
linearity of the stochastic equation (\ref{density1}) in $\theta$, the
number of fields $\theta$ in any 1-irreducible function with the operator
$F(x)$ cannot exceed their
number in the operator itself. This is easily seen from the fact that the
chains of the propagators $\langle\theta'\theta\rangle_{0}$,
$\langle\theta\theta\rangle_{0}$ in any diagram cannot branch; cf. item (iv)
in sec.~\ref{sec:NScomp3}. Then the analysis of expression (\ref{indexo})
shows that the superficial divergence can only be present in the
1-irreducible function with $N_{\theta}=n$ and $N_{\Phi}=0$ for the fields
$\Phi$ other than $\theta$. For this function $\delta_{\Gamma}=0$, the
divergence is logarithmic, and the corresponding counterterm has the form
$\theta^{n}(x)$. Hence, our operators are multiplicatively renormalizable:
$F(x) = Z_{n} F^{R} (x)$ with certain renormalization constants
of the form (\ref{ZMS}).

Now we turn to the calculation of the constants $Z_{n}$ in the leading
(one-loop) approximation.
Let $\Gamma(x;\theta)$ be the generating functional of the
1-irreducible Green functions with one composite operator $F(x)$
and any number of fields $\theta$. Here $x = \{ t,{\bf x}\}$ is
the argument of the operator and $\theta$ is
the functional argument, the ``classical analog'' of the random
field $\theta$. We are interested in the $\theta^{n}$ term of the
expansion of $\Gamma(x;\theta)$ in $\theta(x)$, which we denote
$\Gamma_{n}(x;\theta)$. It can be written as
\begin{widetext}
\begin{equation}
\Gamma_{n}(x;\theta) =  \int dx_{1} \cdots \int dx_{n}
\, \theta(x_{1})\cdots\theta(x_{n})\,
\langle F(x) \theta(x_{1})\cdots\theta(x_{n})\rangle_{\rm 1-ir}.
\label{Gamma1}
\end{equation}
\end{widetext}
In the one-loop approximation the function (\ref{Gamma1}) is
represented diagramatically as follows:
\begin{equation}
\Gamma_{n}(x;\theta)= F(x) + \frac{1}{2}  \vcenter{\hbox
{\includegraphics [width=.09\textwidth,clip]{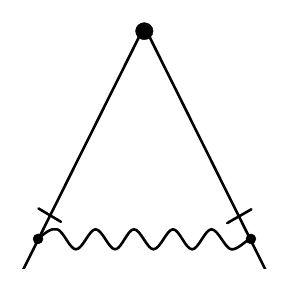}}}
\label{Gamma2}
\end{equation}
The first term is the tree (loopless) approximation, and the thick dot
with the two attached lines in the diagram denotes the operator vertex,
that is, the variational derivative
\begin{equation}
V(x;x_{1},x_{2})=\delta^{2}F(x)/{\delta\theta(x_{1})\delta\theta(x_{2})}.
\label{Vae}
\end{equation}
In the present case, the vertex
\begin{equation}
V(x;x_{1},x_{2}) = n(n-1)\, \theta^{n-2}(x)\, \delta(x-x_{1})\delta(x-x_{2})
\label{VertOp}
\end{equation}
contains $(n-2)$ fields $\theta$. (We recall that
$\delta\theta(x)/\delta\theta(x') = \delta(x-x') \equiv \delta(t-t')
\delta({\bf x}-{\bf x}')$.) Two more fields are attached to the
plain vertices $\theta'\partial(v\theta)$ at the bottom of the diagram.

Since the divergence is logarithmic, one can set all the external
frequencies and momenta equal to zero. Then all $\theta$'s acquire the
common argument $x$ and the diagram becomes proportional to the operator
$\theta^{n}(x)$ with the coefficient, given by the ``core'' of the diagram:
\begin{equation}
\int \frac{d\omega}{2\pi} \int \frac{d{\bf k}}{(2\pi)^{d}} \, k_{s}k_{l}\,
\frac{1}{\omega^{2}+w^{2}\nu^2 k^{4}}\, D_{sl}(\omega, {\bf k}),
\label{Dia11}
\end{equation}
where the first factor in the integrand comes from the vertices
(\ref{vertex1}), the second one comes from the propagators
$\langle\theta'\theta\rangle_{0}$ in (\ref{lines3}) with the replacement
$\kappa_{0} \to w\nu$, and the last factor is the velocity propagator
from (\ref{Self2}). Note that only the second term from $D_{sl}$ gives
nonvanishing contribution to (\ref{Dia11}). Integration over the frequency
is easily performed using the formula
\begin{equation}
\int \frac{d\omega}{2\pi} \frac{1}{(\omega^{2}+a^{2})(\omega^{2}+b^{2})}
= \frac{1}{2ab(a+b)},
\label{Res}
\end{equation}
and after the contraction of the tensor indices the integral over the
momentum reduces to (\ref{ska}). Collecting all the factors gives
\begin{widetext}
\begin{equation}
\Gamma_{n}(x;\theta)= \theta^{n}(x) \, \left\{ 1 +
\frac{n(n-1)}{2}\, \frac{\alpha\hat g}{2wu(u+w)} \,
\left(\frac{\mu}{m}\right)^{y}\, \frac{1}{y} \right\} ,
\label{Gamma22}
\end{equation}
\end{widetext}
with $\hat g$ defined in (\ref{ghat}) and up to a finite part and
higher-order corrections.

The renormalization constant $Z_{n}$ is found from the requirement that the
renormalized analog $\Gamma_{n}^{R}=Z_{n}^{-1}\Gamma_{n}$ of the function
(\ref{Gamma1}) be UV finite in terms of renormalized parameters (mind the
minus sign in the exponent). In our approximation, it is sufficient to
replace $\theta^{n}\to Z_{n}^{-1} \theta^{n}$ only in the first term of the
expression (\ref{Gamma22}) and then to choose $Z_{n}$ to cancel the pole
in the second term. In the MS scheme this gives
\begin{equation}
Z_{n} = 1 + \frac{n(n-1)}{2}\, \frac{\alpha\hat g}{2wu(u+w)} \,
\frac{1}{y}.
\label{Zn}
\end{equation}
Then for the corresponding anomalous dimension eq. (\ref{E}) gives
\begin{equation}
\gamma_{n} = - \frac{n(n-1)}{2}\, \frac{\alpha\hat g}{2wu(u+w)},
\label{gaman}
\end{equation}
with the higher-order corrections in $\hat g$.

For the critical dimensions of the operators $\theta^{n}$ from the
expression (\ref{Krit}) one obtains
\begin{equation}
\Delta[\theta^{n}]= n\Delta_{\theta} + \gamma_{n}^{*},
\label{KrOp}
\end{equation}
and substituting the fixed-point values (\ref{FP}) and $w_{*}=1$ into
(\ref{gaman}) finally gives
\begin{equation}
\Delta[\theta^{n}]= -n + \frac{ny}{6} - \frac{n(n-1)\,\alpha\, dy}{6(d-1)},
\label{KrOp1}
\end{equation}
with the higher-order corrections in $y$.
These dimensions are negative (``dangerous'' in the terminology of
\cite{Book3,JETP,UFN,turbo})
and decrease as $n$ grows. One can argue that dangerous operators can always
appear in a field theoretic only as infinite families with the spectrum of
dimensions not bounded from below.

Now let us turn to the same operators $\theta^{n}$ in the tracer model. From
the expression (\ref{indexo}) and the linearity of the stochastic equation
(\ref{tracer1}) it follows that, like for the density case, the superficial
UV divergences can only be present in the 1-irreducible function
$\langle \theta^{n}(x) \theta(x_{1})\cdots\theta(x_{n})\rangle_{\rm 1-ir}$.
Clearly, at least one of the external tails of the field $\theta$ is attached
to a vertex $\theta'(v\partial)\theta$: it is impossible to construct a
nontrivial diagram of the desired type with all the external tails attached
only to the vertex (\ref{VertOp}) of the operator $F(x)$. Therefore at least
one derivative $\partial$, acting on a tail $\theta$, appears as an external
factor in the diagram. Consequently, its real index of divergence
$\delta_{\Gamma}'$ is necessarily negative, and the diagram is in fact UV
convergent; cf. item (iii) in sec.~\ref{sec:NScomp3}.

This means that the operators $\theta^{n}$ are in fact UV finite, $Z_{n}=1$,
and their scaling dimensions are given by the expression
\begin{equation}
\Delta[\theta^{n}]= n\Delta_{\theta} = -n + {ny}/{6}
\label{KrOp2}
\end{equation}
exactly, that is, with no higher-order corrections in $y$.

\subsection{Renormalization of the composite fields $(\partial\theta)^{n}$
in the tracer model. Explicit leading-order results} \label{sec:Polly}

In the tracer model, of special importance are tensor operators, constructed
solely of the gradients of the passive scalar field. Such operators with the
lowest canonical dimension contain the minimal number of derivatives (one
derivative per each field) and have the form
\begin{equation}
F^{(n,l)}_{i_{1}\dots i_{l}} =
\partial_{i_{1}}\theta\cdots\partial_{i_{l}}\theta\,
(\partial_{i}\theta\partial_{i}\theta)^{s} + \dots.
\label{Fnp}
\end{equation}
Here $l$ is the number of the free vector indices (the rank of the tensor)
and $n=l+2s$ is the total number of the fields $\theta$ entering into the
operator. The ellipsis stands for the subtractions with Kronecker's delta
symbols that make the operator irreducible (so that contraction with respect
to any pair of the free tensor indices vanish), for example,
\begin{equation}
F^{(2,2)}_{ij} = \partial_{i}\theta \partial_{j}\theta -
\frac{\delta_{ij}}{d}\, (\partial_{k}\theta\partial_{k}\theta).
\label{Irr}
\end{equation}

For all these operators $d_{F}=0$, and the real index of divergence
is $\delta'_{\Gamma} = \delta_{\Gamma} - N_{\theta}$ with $\delta_{\Gamma}$
from (\ref{indexo}). Indeed, now one derivative $\partial$ appears as an
external factor in a diagram for any external tail $\theta$, no matter is
it attached to the ordinary vertex $\theta'(v\partial)\theta$ or to the
vertex (\ref{VertOp}) for the operator (\ref{Fnp}). Like for the operators
$\theta^{n}$, the number of the fields $\theta$ in any 1-irreducible function
cannot exceed their number in the operator itself: $N_{\theta} \le n$, cf.
the discussion in sec.~\ref{sec:CF}. It then follows that superficial UV
divergences can only be present in the 1-irreducible functions
$\langle F^{(n,l)}(x) \theta(x_{1}) \dots \theta(x_{k}) \rangle_{1-ir}$
with $k \le n$. For such functions $\delta'_{\Gamma}=0$ and
$\delta_{\Gamma}=k$, so that the corresponding counterterm can only
involve the monomials $F^{(k,p)}$ from (\ref{Fnp}) with certain values
of the rank $p$. We conclude that the family of the operators (\ref{Fnp})
is closed with respect to renormalization in the sense that
$F^{(n,l)} = Z_{(n,l)(k,p)} F^{(k,p)}_{R}$ with a certain matrix of
renormalization constants. Since $Z_{(n,l)(k,p)}=0$ for $k>n$, this matrix
is block-triangular with the diagonal sub-blocks corresponding to $n=k$,
and so is the corresponding matrix $\Delta_{F}$ in (\ref{Krit}).

We are interested presumably in the scaling dimensions, associated with the
operators (\ref{Fnp}). They are given by the eigenvalues of the matrix
$\Delta_{F}$, which are completely defined by its diagonal sub-blocks.
A simple analysis shows that the corresponding diagrams do not involve the
propagator $\langle\theta\theta\rangle_{0}$ from (\ref{lines3}); this is
again a consequence of the linearity of the original stochastic equation
(\ref{tracer1}). Hence, the diagonal blocks can be calculated directly in
the model without the random noise in (\ref{tracer1}), because the
correlation function of the noise (\ref{noise}) enters the diagrams only
via the propagator $\langle\theta\theta\rangle_{0}$. The function
(\ref{noise}) is the only source of the anisotropy in the problem.
Without the noise, the model becomes $SO(d)$ covariant, and the
irreducible tensor
operators with different ranks cannot mix in renormalization. This means
that the diagonal sub-blocks of the matrix $\Delta_{F}$ are in fact
diagonal, and their diagonal elements coincide with the eigenvalues of
the full matrix $\Delta_{F}$.

We finally conclude that, as long as the scaling dimensions are concerned,
the operators (\ref{Fnp}) can be treated as multiplicatively renormalizable,
$F^{(n,l)} = Z_{(n,l)} F^{(n,l)}_{R}$ with certain renormalization
constants $Z_{(n,l)}$, the diagonal elements of the full matrix
$Z_{(n,l)(k,p)}$.

For practical calculations, it is convenient to contract the tensors
(\ref{Fnp}) with an arbitrary constant vector
{\mbox{\boldmath $\lambda$}}$=\{\lambda_{i}\}$.
The resulting scalar operator has the form
\begin{equation}
F^{(n,l)} = (\lambda_{i}w_{i})^{l} (w_{i}w_{i})^{s} + \dots,
\quad w_{i} \equiv \partial_{i}\theta,
\label{FnpSk}
\end{equation}
where the subtractions, denoted by the ellipsis, necessarily involve the
factors of $\lambda^{2}=\lambda_{i}\lambda_{i}$.
The counterterm to $F^{(n,l)}$ is proportional to the same operator,
and in order to find the constant $Z_{(n,l)}$, it is sufficient to
retain only the principal monomial, explicitly shown in (\ref{FnpSk}),
and to discard in the result all the terms with factors of $\lambda^{2}$.
Then, using the chain rule, the vertex (\ref{Vae}) for the operator
$F^{(n,l)}$ can be written in the form
\begin{equation}
V(x;x_{1},x_{2})=  \frac{\partial^{2}F^{(n,l)}}{\partial w_{i}\partial w_{j}}
\,\partial_{i}\delta(x-x_{1})\, \partial_{j}\delta(x-x_{2})
\label{Vae1}
\end{equation}
up to irrelevant terms. The differentiation gives
\begin{widetext}
\begin{eqnarray}
{\partial^{2}F^{(n,l)}}/{\partial w_{i}\partial w_{j}} &=&
2s (w^{2})^{s-2} (\lambda w)^{l} \left[\delta_{ij} w^{2} +2(s-1)w_{i}w_{j}
\right] + l(l-1) (w^{2})^{s}(\lambda w)^{l-2} \lambda_{i} \lambda_{j}+
\nonumber \\
&+& 2ls (w^{2})^{s-1}(\lambda w)^{l-1} (w_{i}\lambda_{j}+ w_{j}\lambda_{i}),
\label{Vae11}
\end{eqnarray}
\end{widetext}
where $w^{2}=w_{k}w_{k}$ and $(\lambda w)=\lambda_{k}w_{k}$. Two more
factors $w_{p}w_{r}$ are attached to the bottom of the diagram, the
derivatives coming from the vertices $\theta'(v\partial)\theta$. The UV
divergence is logarithmic, and one can set all the external frequency and
momentum equal to zero; then the core of the diagram takes on the form
\begin{equation}
\int\frac{d\omega}{2\pi}
\int_{k>m}\frac{d{\bf k}}{(2\pi)^{d}}\,
k_{i}k_{j}\, D_{pr} (\omega, {\bf k})\,
\frac{1}{\omega^{2}+w^{2}\nu^2 k^{4}}.
\label{triad}
\end{equation}
Here the first factor comes from the derivatives in (\ref{Vae1}), $D_{pr}$
from (\ref{Self2}) is the velocity correlation function (\ref{lines}), and
the last
factor comes from the two propagators $\langle\theta'\theta\rangle_{0}$.
The substitutions $Z\to1$, $c\to0$ are made; cf. the discussion in
sec.~\ref{sec:PD3}.

Intergations over the frequency are easily performed using (\ref{Res}),
then all the resulting integrals over ${\bf k}$ are reduced to the scalar
integral (\ref{ska}) using the relations (\ref{tenz}). Combining all the
factors, contracting the tensor indices and expressing the result in
$n=l+2s$ and $l$ gives:
\begin{widetext}
\begin{equation}
\Gamma_{n}(x;\theta)= F^{(n,l)}(x)\, \left\{ 1 - \frac{\hat g}{2yd(d+2)}\,
\left(\frac{\mu}{m}\right)^{y}\,
\left(
\frac{Q_{1}}{2w(1+w)} + \alpha  \frac{Q_{2}}{2wu(u+w)}
\right) \right\} ,
\label{QQ}
\end{equation}
where
\begin{eqnarray}
Q_{1} = -n(n+d)(d-1) + (d+1) l(l+d-2), \qquad
Q_{2} = -n(3n+d-4) + l(l+d-2),
\label{Q}
\end{eqnarray}
\end{widetext}
and $\hat g$ is defined in (\ref{ghat}). Then the renormalization constant
$Z_{(n,l)}$ in the MS scheme reads
\begin{equation}
Z_{(n,l)} = 1 - \frac{\hat g}{2yd(d+2)}\, \left\{
\frac{Q_{1}}{2w(1+w)} + \alpha  \frac{Q_{1}}{2wu(u+w)}
\right\},
\label{ZZ}
\end{equation}
see the explanation in sec.~\ref{sec:CF} below eq.~(\ref{Gamma22}).
Then for the corresponding anomalous dimension eq.~(\ref{E}) gives
\begin{equation}
\gamma_{(n,l)} = \frac{\hat g}{2d(d+2)}\, \left\{
\frac{Q_{1}}{2w(1+w)} + \alpha  \frac{Q_{1}}{2wu(u+w)}
\right\},
\label{GG}
\end{equation}
with the higher-order corrections in $\hat g$.

Finally, for the scaling dimension, associated with the operators
(\ref{Fnp}), the general expression (\ref{Krit}) gives
\begin{equation}
\Delta_{(n,l)} = n + n\Delta_{\theta} + \gamma_{(n,l)}^{*} =
ny/6+ \gamma_{(n,l)}^{*}.
\label{Krnl}
\end{equation}
Substituting the fixed-point values (\ref{FP}) and $w_{*}=1$ into
(\ref{GG}) one finally obtains
\begin{equation}
\Delta_{(n,l)} = ny/6+
\frac{y \left\{ {Q_{1}} + \alpha  {Q_{1}} \right\}} {6(d-1)(d+2)},
\label{Dnl}
\end{equation}
with the higher-order corrections in $y$.

In particular, for the scalar operator with $l=0$ one obtains:
\begin{equation}
\Delta_{(n,0)} = \frac{-yn\{(n-2)(d-1)+\alpha (3n+d-4)\}}{6(d-1)(d+2)}.
\label{Dnn}
\end{equation}
Again, we meet
an infinite family of dangerous operators with the spectrum of dimensions
not bounded from below. For a fixed $n$, the dimension (\ref{Dnl})
increases with the rank $l$, so that for the maximum possible rank
$l=n$ one always has $\Delta_{(n,n)}>0$. This hierarchy, which is
conveniently expressed by the inequality $\partial_{l} \Delta_{(n,l)} >0$,
becomes more strongly pronounced when $\alpha$ grows:
$\partial_{l}\partial_{\alpha} \Delta_{(n,l)} >0$. All these properties
will be important in the OPE analysis of sec.~\ref{sec:OPE}.

\subsection{More tensor operators} \label{sec:summa}

We will also need to know the critical dimensions of the $l$th rank
irreducible tensor operators, built only of two fields $\theta$ and $l$
spatial derivatives. An example is provided by the operator
\begin{equation}
F_{i_{1}\dots i_{l}}(x) =  \theta(x) \partial_{i_{1}}
\cdots\partial_{i_{l}} \theta(x) + \dots.
\label{Fl}
\end{equation}
As earlier in (\ref{Fnp}), the ellipsis stands for the subtractions with
Kronecker's delta symbols that make the operator irreducible. Of course,
for any given $l>1$, there are several such operators with different
placement of the derivatives: in the special case (\ref{Fl}), all the
derivatives act on the same field. However, all the other such operators
differ from (\ref{Fl}) by a total derivative, which is easily seen from
the relation
\begin{equation}
F(x) \partial G(x) = - G(x) \partial F(x) + \partial (F(x) G(x)).
\label{Nwt}
\end{equation}
Thus the set of independent
$l$th rank operators can be chosen as (\ref{Fl}) and the operators
having the forms of derivatives, for example, for $l=2$, as
$\theta \partial_{i}\partial_{j} \theta + \dots$ and
$\partial_{i}\partial_{j}(\theta\theta)  + \dots$. In the calculation of
their critical dimensions, it is sufficient to consider the $SO(d)$ covariant
model without the noise (\ref{noise}); see the discussion in the
preceding subsection. Then the operators with different ranks do not mix
in renormalization. The analysis of renormalization also shows that the
operator (\ref{Fl}) can mix only with its own ``family'' of derivatives:
the operators with additional derivatives (like $\partial_{t}$ or
$\partial^{2}$) or with the fields $\theta'$, $\phi$, $\phi'$, $v'$ have
too high canonical dimensions $d_{F}$, the appearance of $v$ is forbidden
by Galilean symmetry, and extra $\theta$'s are forbidden by the linearity
of the model.

The same relation (\ref{Nwt}) also shows that for odd $l$, the operator
(\ref{Fl}) itself reduces to a derivative (more precisely, to a linear
combination of derivatives). In the following, we will be interested only
in the operators not reducible to derivatives, and thus, from now on, we
will consider only
even values of $l$. Then (\ref{Fl}) is nontrivial and it cannot admix to the
derivatives from its family, although they can admix to (\ref{Fl}). Thus the
corresponding renormalization matrix $Z_{F}$ appears block triangular,
and so is the matrix $\Delta_{F}$. The eigenvalue, associated with the
nontrivial operator (\ref{Fl}), coincides with the corresponding diagonal
element of $\Delta_{F}$. We conclude that in the calculation of the critical
dimension, associated with the operator (\ref{Fl}), the latter can be treated
as if it were multiplicatively renormalizable.

Like in the preceding subsection, it is convenient to contract the operator
 (\ref{Fl}) with an arbitrary constant vector
{\mbox{\boldmath $\lambda$}}$=\{\lambda_{i}\}$.
The resulting scalar operator has the form
\begin{equation}
F_{l} =  \theta (\lambda_{i}\partial_{i})^{l} \theta + \dots,
\label{FlSk}
\end{equation}
where the terms, denoted by the ellipsis, necessarily involve the
factors of $\lambda^{2}$. In order to find the corresponding
renormalization constant $Z_l$, it is sufficient to
keep only the principal monomial, explicitly shown in (\ref{FnpSk}), and
to retain in the result for the counterterm only terms of the same form.
Then the relevant part of the vertex factor (\ref{Vae}) is
\begin{equation}
V(x;x_{1},x_{2}) = \delta(x-x_{1}) (\lambda_{i}\partial_{i})^{l}
\delta(x-x_{2}) + \{ x_{1} \leftrightarrow x_{2} \}.
\label{Wae}
\end{equation}

The one-loop approximation for the functional (\ref{Gamma1}) for the
operator (\ref{FlSk}) has the same form (\ref{Gamma2}).
Let us choose the external momentum ${\bf p}$ to flow into the diagram
through the left lower vertex and to flow out through the right lower one.
The external momentum flowing through operator's  vertex and all the
external frequencies are set equal to zero: this is sufficient to find
the needed counterterm. Furthermore, we will put $w=u=1$ in the propagators
from the very beginning, because we are eventually interested in the
value of the anomalous dimension at the fixed point $w_{*}=u_{*}=1$.

Let us begin with the tracer case. Then the core of the diagram in
(\ref{Gamma2}) takes on the form
\begin{widetext}
\begin{equation}
p_{i}p_{j}\,  \int \frac{d\omega}{2\pi} \int_{k>m}
\frac{d{\bf k}}{(2\pi)^{d}} \,
2{\rm i}^{l} ( {\mbox{\boldmath $\lambda$}}  {\bf q})^{l} \,
\frac{g\mu^y \nu^3 k^{4-d-y}}{\omega^{2}+ \nu^2 k^{4}}\,
\left\{ P^{\bot}_{ij} ({\bf k})
+ \alpha P^{\parallel}_{ij} ({\bf k}) \right\}\,
\frac{1}{\omega^{2}+ \nu^2 q^{4}}.
\label{Hard}
\end{equation}
\end{widetext}
Here the factor $p_{i}p_{j}$ comes from the vertices (\ref{vertex2}), the
factor $2{\rm i}^{l}$({\mbox{\boldmath $\lambda$}${\bf q})^{l}$ comes from
the vertex (\ref{Wae}) for even $l$ (for the odd $l$ the two terms in
(\ref{Wae}) would cancel each other and instead of factor 2 one would
get 0), the factors depending on ${\bf k}$ represent
the velocity correlation function from
(\ref{lines}) with the proper substitutions, including $c=0$ and $w=u=1$.
The last factor comes from the propagators $\langle\theta'\theta\rangle_{0}$.
The momentum ${\bf k}$ flows through the velocity propagator, so that
${\bf q}= {\bf k}+{\bf p}$.

In the resulting expression we retain only terms of the form
({\mbox{\boldmath $\lambda$}}${\bf p})^{l}$ and drop all the other terms,
containing $\lambda^{2}$ or $p^{2}$. Thus we can replace
\[ p_ip_{j}\left\{P^{\bot}_{ij}+ \alpha P^{\parallel}_{ij}\right\}
\to  (\alpha-1) ({\bf pk})^{2} /k^{2}. \]
The integration over $\omega$ in (\ref{Hard}) is easily
performed using (\ref{Res}) and gives:
\begin{equation}
g \mu^y (\alpha-1){\rm i}^{l}
\int_{k>m} \frac{d{\bf k}}{(2\pi)^{d}} \,  ({\bf pk})^{2}
( {\mbox{\boldmath $\lambda$}}  {\bf q})^{l} \,
\frac{k^{-d-y}}{q^{2}(k^{2}+q^{2})}.
\label{ioo}
\end{equation}

Now we expand all the denominators in the integrand of (\ref{Hard})
in ${\bf p}$ (dropping all the terms with $p^{2}$):
\begin{widetext}
\begin{eqnarray}
\frac{1}{q^{2}} \simeq \frac{1}{k^{2}+2({\bf pk})} = \frac{1}{k^{2}}
\sum^{\infty}_{s=0} \frac{(-2)^{s}({\bf pk})^{s}}{k^{2s}},
\quad
\frac{1}{k^{2}+q^{2}} \simeq \frac{1}{2(k^{2}+{\bf pk})} = \frac{1}{2k^{2}}
\sum^{\infty}_{m=0} \frac{(-1)^{m}({\bf pk})^{m}}{k^{2m}},
\label{Tay}
\end{eqnarray}
\end{widetext}
and expand the numerator using Newton's binomial formula:
\begin{eqnarray}
({\mbox{\boldmath $\lambda$}}{\bf q})^{l} = \sum_{n=0}^{l}
C_{l}^{n} ({\mbox{\boldmath $\lambda$}}{\bf k})^{n}
({\mbox{\boldmath $\lambda$}}{\bf p})^{l-n}.
\label{Bin}
\end{eqnarray}

In the resulting three-fold series over $n,m,s$
\[ \sum_{n=0}^{l} C_{l}^{n} ({\mbox{\boldmath $\lambda$}}{\bf p})^{l-n}
\sum_{m,s=0}^{\infty} \frac{(-1)^{m} (-2)^{s}
({\bf pk})^{m+s+2} ({\mbox{\boldmath $\lambda$}}{\bf k})^{n}}
{k^{2(s+m)}} \]
we only need to collect the terms proportional to
({\mbox{\boldmath $\lambda$}}${\bf p})^{l}$, which leads to the
restriction $n=s+m+2$ and hence to the finite double sum:
\begin{widetext}
\begin{eqnarray}
\sum_{s,m=0}^{s+m+2\le l} (-1)^{m} (-2)^{s} C_{l}^{s+m+2}
\frac{({\mbox{\boldmath $\lambda$}}{\bf p})^{l-m-s-2}
({\bf pk})^{m+s+2} ({\mbox{\boldmath $\lambda$}}{\bf k})^{s+m+2}}
{k^{2(s+m)}}.
\label{sum1}
\end{eqnarray}
\end{widetext}

Substituting it to the (\ref{ioo}) gives rise to the integrals
\begin{eqnarray}
J_{i_{1} \dots i_{2n}} (m) =
\int_{k>m} \frac{d{\bf k}}{(2\pi)^{d}} \, k^{-d-y} \,
\frac{k_{i_{1}} \dots k_{i_{2n}}}    {k^{2n}}
\label{Ints}
\end{eqnarray}
with $n=s+m+2 \ge 2$.
They are easily found using the isotropy considerations, cf. (\ref{tenz}):
\begin{eqnarray}
J_{i_{1} \dots i_{2n}} (m) = \frac{\delta_{i_{1}i_{2}}\dots
\delta_{i_{2n-1}i_{2n}} + {\rm all\ permutations}}
{d(d+2)\dots (d+2n-2)}\, J(m)
\nonumber \\ {}
\label{Jnt}
\end{eqnarray}
with $J(m)$ from (\ref{ska}). The sum over all possible permutations of
$2n$ tensor indices in the numerator of (\ref{Jnt}) involves
$(2n-1)!! = (2n)!/2^{n}n!$ terms, but we have to keep only the terms
that give rise to the structure
({\mbox{\boldmath $\lambda$}}${\bf p})^{n}$
after the contraction with the vectors {\mbox{\boldmath $\lambda$}} and
${\bf p}$ in (\ref{sum1}). It is easy to grasp that there are only
$n!$ such permutations.

Collecting all the factors gives for the core (\ref{Hard}) of the diagram
in (\ref{Gamma2}) the following expression:
\begin{eqnarray}
{\rm i}^{l} ({\mbox{\boldmath $\lambda$}}{\bf p})^{l}
(\alpha-1) {\hat g} \left(\frac{\mu}{m}\right)^{y} \frac{1}{2y}\,
{\cal S}_{l}(d),
\label{FiR}
\end{eqnarray}
where $\hat g$ is defined in (\ref{ghat}) and
\begin{eqnarray}
{\cal S}_{l}(d) = \sum_{s,m=0}^{s+m+2\le l}
\frac{(-1)^{s+m}2^{s} C_{l}^{s+m+2}(s+m+2)!}
{d(d+2)\dots (d+2(s+m)+2)}.
\nonumber \\ {}
\label{Sl}
\end{eqnarray}
For $l=0$, the sums (\ref{sum1}) and (\ref{Sl})  contain no terms, so that
${\cal S}_{0}(d) = 0$.

For the functional (\ref{Gamma1}) we then obtain (i$p_{i}\to\partial_{i}$)
\begin{eqnarray}
\Gamma_{2}(x) = F_{l}(x) \left\{1 +
(\alpha-1) \frac{{\hat g}}{4y}  \left(\frac{\mu}{m}\right)^{y}\,
{\cal S}_{l}(d)
\right\}
\nonumber \\ {}
\label{Pf}
\end{eqnarray}
with the operator $F_{l}$ from (\ref{FlSk}); note the additional factor $1/2$
from the symmetry coefficient in (\ref{Gamma1}). Then for the
renormalization constant from the relation $F_{l}=Z_{l}F_{l}^{R}$ in the
MS scheme we obtain
\begin{eqnarray}
Z_{l}= 1 + (\alpha-1)  \frac{{\hat g}}{4y}\, {\cal S}_{l}(d) ,
\label{Zl}
\end{eqnarray}
and the corresponding anomalous dimension is:
\begin{eqnarray}
\gamma_{l}(g)= - (\alpha-1)  \frac{{\hat g}}{4}\, {\cal S}_{l}(d).
\label{gaml}
\end{eqnarray}

The sum ${\cal S}_{l}(d)$ in (\ref{Sl})  can be reduced to a simpler
one-fold sum for general $l$. Let us pass from $s$ and $m$ to the new
summation variables $k=s+m$ and $m$ and substitute the explicit expression
for the binomial coefficient $C^{k+2}_{l} = l!/(k+2)!(l-k-2)!$. This gives
\begin{widetext}
\begin{eqnarray}
{\cal S}_{l}(d)= l!\,
\sum_{k=0}^{k+2\le l} \left\{ \sum_{m=0}^{k} \frac{1}{2^{m}} \right\}\,
\frac{(-2)^{k}} {(l-k-2)!\,d(d+2)\dots (d+2k+2)}.
\label{sum2}
\end{eqnarray}
\end{widetext}
Now the internal summation over $m$ is readily performed to give
$2-2^{-k}$, so that, after changing the summation variable $k\to k+2$,
we obtain
\begin{eqnarray}
{\cal S}_{l}(d)= 2{\cal N}_{l}(d) -{\cal M}_{l}(d),
\label{sum31}
\end{eqnarray}
where
\begin{eqnarray}
{\cal N}_{l}(d)= l!\,\sum_{k=2}^{l}
\frac{(-2)^{k-2}}{(l-k)!\,d(d+2)\dots (d+2k-2)},
\nonumber \\
{\cal M}_{l}(d)= l!\, \sum_{k=2}^{l}
\frac{(-1)^{k-2}}{(l-k)!\,d(d+2)\dots (d+2k-2)}.
\nonumber \\ {}
\label{sum3}
\end{eqnarray}
The first sum can be calculated explicitly for any $l$, cf. \cite{AG}:
\begin{eqnarray}
{\cal N}_{l}(d)= \frac{4l(l-1)}{d(d+2l-2)},
\label{sum4}
\end{eqnarray}
while the second can be easily calculated for any given $l$.

For the critical dimension, associated with the operator (\ref{Fl}),
from the relation (\ref{Krit}) we finally obtain:
\begin{eqnarray}
\Delta_{l} = l+2\Delta_{\theta}+\gamma_{l}^{*}= l-2+y/3+ \gamma_{l}^{*},
\label{Dl}
\end{eqnarray}
where from (\ref{gaml}) and (\ref{FP}) we find
\begin{eqnarray}
\gamma_{l}^{*}=\gamma_{l}(g^{*}) = -   \frac{yd(\alpha-1)}{3(d-1)}\,
{\cal S}_{l}(d),
\label{gamZ}
\end{eqnarray}
with the higher-order corrections in $y$.

For $l=0$, expressions (\ref{gaml}), (\ref{Dl}) agree with the exact result
(\ref{KrOp2}) for the operator $\theta^{2}$ (we recall that
${\cal S}_{0}(d)=0$), while for $l=2$ they agree with the results
(\ref{GG})--(\ref{Dnl}) with $n=l=2$.

Now let us turn to the density case. Then the factor $p_{i}p_{j}$ in
(\ref{Hard}) should be replaced with $q_{i}q_{j}$ (and, of course, moved
into the integrand). It is convenient to write
\begin{widetext}
\begin{equation}
q_{i}q_{j}\, \left\{ P^{\bot}_{ij} ({\bf k})
+ \alpha P^{\parallel}_{ij} ({\bf k}) \right\} =
p_{i}p_{j} \left\{ P^{\bot}_{ij} ({\bf k})
+ \alpha P^{\parallel}_{ij} ({\bf k}) \right\} + \alpha (q^{2}-p^{2}):
\label{Moo}
\end{equation}
\end{widetext}
the first term gives the old expression (\ref{Hard}), and the last one is
proportional to $p^{2}$ and can be dropped. Thus we only need to calculate
the contribution of the term $\alpha q^{2}$ to the analog of expression
(\ref{Hard}). Then the analog of (\ref{ioo}) takes on the form
\begin{eqnarray}
g \mu^y \alpha\, {\rm i}^{l} \int_{k>m} \frac{d{\bf k}}{(2\pi)^{d}} \,
( {\mbox{\boldmath $\lambda$}}  {\bf q})^{l} \,
\frac{k^{2-d-y}}{(k^{2}+q^{2})}.
\label{ana2}
\end{eqnarray}
Applying the expansions (\ref{Tay}), (\ref{Bin}) leads to the double sum
\[ \sum_{n=0}^{l} C_{l}^{n} ({\mbox{\boldmath $\lambda$}}{\bf p})^{l-n}
\sum_{m=0}^{\infty} \frac{(-1)^{m}
({\bf pk})^{m} ({\mbox{\boldmath $\lambda$}}{\bf k})^{n}}
{k^{2m}}. \]
We have to retain only the terms proportional to
({\mbox{\boldmath $\lambda$}}${\bf p})^{l}$, which leads to the
restriction $n=m$ and hence to the finite sum:
\begin{eqnarray}
\sum_{m=0}^{l} (-1)^{m}  C_{l}^{m}
\frac{({\mbox{\boldmath $\lambda$}}{\bf p})^{l-m}
({\bf pk})^{m} ({\mbox{\boldmath $\lambda$}}{\bf k})^{m}}
{k^{2m}}.
\label{ana3}
\end{eqnarray}
Substituting it into (\ref{ana2}) gives rise to the integrals (\ref{Ints})
with all $n\ge0$. In the sum (\ref{Jnt}) over all possible permutations
we have to keep only $n!=m!$ terms that give rise to the structure
({\mbox{\boldmath $\lambda$}}${\bf p})^{n}$ after the contraction with
the vectors {\mbox{\boldmath $\lambda$}} and ${\bf p}$ in (\ref{ana3}).
To avoid possible confusion, we will write down the terms with $m=0$ and
$m=1$ separately and for $m\ge2$ apply the formula (\ref{Jnt}). Then
collecting all terms gives the following result for (\ref{ana2}):
\begin{eqnarray}
{\rm i}^{l} ({\mbox{\boldmath $\lambda$}}{\bf p})^{m}
\frac{g\mu^y\alpha}{2}\, \left\{ 1-\frac{l}{d}+{\cal M}_{l}(d)
\right\}\, J(m)
\label{ana4}
\end{eqnarray}
with $J(m)$ from (\ref{ska}) and the sum ${\cal M}_{l}(d)$ from (\ref{sum3}).

Proceeding as before for the tracer case, we arrive at the following
expression for the renormalization constant $Z_{l}$ in the MS scheme:
\begin{eqnarray}
Z_{l} = 1 +(\alpha-1)\frac{\hat g}{4y} {\cal S}_{l}(d) +
\alpha \frac{\hat g}{4y} \left\{ 1-\frac{l}{d}+{\cal M}_{l}(d)
\right\}.
\nonumber \\ {}
\label{ana5}
\end{eqnarray}
Here the contribution with ${\cal S}_{l}(d)$ comes from the first term in
(\ref{Moo}) and the last term with curly brackets comes from (\ref{ana5}).
Then for the anomalous dimension, using the expressions
(\ref{sum31})--(\ref{sum4}), we obtain
\begin{eqnarray}
\gamma_{l}(g) = -\alpha \frac{\hat g}{4} \left(1-\frac{l}{d}\right) +
(1-\alpha) \frac{\hat g}{4} {\cal N}_{l}(d)
 - \frac{\hat g}{4}{\cal M}_{l}(d),
\nonumber \\ {}
\label{ana6}
\end{eqnarray}
with higher-order corrections in $g$.

In the expression (\ref{Dl}) for the critical dimension one has:
\begin{widetext}
\begin{eqnarray}
\gamma_{l}^{*} = -\alpha\frac{y(l-d)}{3(d-1)} + (1-\alpha)
\frac{8l(l-1)y}{3(d-1)(d+2l-2)} - \frac{dy}{3(d-1)} {\cal M}_{l}(d),
\label{ana7}
\end{eqnarray}
\end{widetext}
with higher-order corrections in $y$. For $l=0$ this result is in agreement
with the expression (\ref{KrOp1}) for the operator $\theta^{2}$ in the
density case.

\section{Operator product expansion and the anomalous scaling}\label{sec:OPE}

\subsection{The case of a density field} \label{sec:OPED}

Consider the equal-time pair correlation function of two UV finite quantities
$F_{1,2}(x)$ with definite critical dimensions, for example, those
of the primary
fields or renormalized local composite operators. We restrict ourselves with
equal-time correlators, because they are usually Galilean invariant and do
not bear strong dependence on the IR scale, caused by the so-called sweeping
effects. From the (canonical) dimensionality considerations it follows that
\begin{eqnarray}
\langle F_{1}(t,{\bf x}_{1}) F_{2}(t,{\bf x}_{2}) \rangle =
\nu^{d^{\omega}_{F}} \mu^{d_{F}} \eta(\mu r, mr, c/(\mu\nu)),
\nonumber \\  {}
\label{Parr0}
\end{eqnarray}
where $d^{\omega}_{F}$ and $d_{F}$ are the canonical dimensions of the
correlation function, given by simple sums of the corresponding dimensions
of the operators, $r=|{\bf x}_{2}-{\bf x}_{1}|$,
and $\eta(\dots)$ is a function of completely dimensionless
variables. We have written the right hand side in terms of renormalized
parameters, when the reference mass substitutes the typical UV momentum
scale $\Lambda$. The behavior of the function $\eta$ in the IR range, that
is, for $\mu r \gg1$, is determined by the IR attractive fixed point of the
RG equation. Solving the RG equation in a standard way, one derives the
following asymptotic expression:
\begin{equation}
\langle F_{1}(t,{\bf x}_{1}) F_{2}(t,{\bf x}_{2}) \rangle \simeq
\nu^{d^{\omega}_{F}} \mu^{d_{F}}(\mu r)^{-\Delta_{F}}
\zeta(mr, c(r)).
\label{Parr}
\end{equation}
Here $\Delta_{F}$ is the critical dimension of the correlation function,
given by simple sum of the dimensions of the operators. The RG equation
does not determine the form of the scaling function $\zeta$; it only
determines the form of its arguments. They are canonically and critically
dimensionless: in particular,
\begin{equation}
c(r) = c (\mu r)^{\Delta_{c}} / (\mu\nu)
\label{rms}
\end{equation}
with $\Delta_{c}$ from (\ref{Krit2}) can be interpreted as effective speed of
sound; more detailed discussion of this point can be found in \cite{ANU}.

For the correlation functions of two operators of the type $\theta^{n}(x)$
the general expression (\ref{Parr}) gives:
\begin{equation}
\langle \theta^{p}(t,{\bf x}_{1}) \theta^{k}(t,{\bf x}_{2}) \rangle \simeq
\mu^{-(p+k)} (\mu r)^{-\Delta_{p}-\Delta_{k}} \zeta_{pk}(mr,c(r))
\label{Parr2}
\end{equation}
with the dimensions $\Delta_{n}$ from (\ref{KrOp1}). In the following, we
do not display the dependence on the UV parameters $\mu$ and $\nu$
and omit the indices of the scaling functions.

The inertial-convective range corresponds to the additional condition that
$mr\ll1$. The behavior of the functions $\zeta$ at $mr\to0$ can be
studied by means of the operator product expansion \cite{Zinn,Book3}.
In the case at hand it has the form
\begin{equation}
F_{1}(t,{\bf x}_{1}) F_{2}(t,{\bf x}_{2}) \simeq \sum_{F} C_{F}(mr, c(r))\,
F(t,{\bf x}),
\label{OPE}
\end{equation}
where ${\bf x}_{2}-{\bf x}_{1}\to 0$ and
${\bf x}=({\bf x}_{1}+{\bf x}_{1})/2$
is fixed. The summation in (\ref{OPE}) is taken, in general, over all
possible renormalized local composite operators allowed by the symmetries
of the model and of the left hand side, $C_{F}$ being numerical coefficient
functions analytical in $mr$ and $c(r)$. In our model, due to the linearity
in the field $\theta$, the number of such fields in the operators $F$
cannot exceed their number on the left hand side. This restriction, which
our model shares with the Kraichnan's model and its relatives \cite{RG}
will be very important in the following.

The correlation function (\ref{Parr}) is obtained by averaging (\ref{OPE})
with the weight $\exp{\cal S}_{R}$ with the renormalized action functional
from (\ref{Fact}). The mean values $\langle F(x) \rangle$ appear on the
right hand side. Without loss of generality, it can be assumed that the
expansion in (\ref{OPE}) is made in irreducible tensor operators.
Then, if the model is $SO(d)$ covariant (the correlation function of the
scalar noise (\ref{noise}) depends only on $r=|{\bf r}|$), only scalar
operators survive the averaging. It can also be assumed that the expansion
is made in the operators with definite critical dimensions. Then their
mean values, in the asymptotic region of small $m$, take on the forms
\begin{equation}
\langle F(x) \rangle \simeq m^{\Delta_{F}} \xi(c(1/m)),
\label{Mean}
\end{equation}
with another set of scaling functions $\xi$ and the argument $c(\dots)$
from (\ref{rms}). Since the diagrams of the perturbation theory have finite
limits both for $c\to\infty$ and $c\to 0$, we may assume that the functions
$\xi$ are restricted for all
values of $c$ and can be estimated by some constants. What is more, for $y$
large enough, including the most realistic case $y\to4$, the dimension
$\Delta_{c}$ becomes negative; see expression (\ref{Krit2}). Thus the
argument $c(1/m) \sim c m^{-\Delta_{c}}$ becomes small for fixed $c$ and
$m\to0$, and the function $\xi$ can be replaced by its (finite) limit value
$\xi(0)$. We finally conclude that, in the IR range,
\begin{equation}
\langle F(x) \rangle \sim m^{\Delta_{F}}.
\label{Mean2}
\end{equation}
Then combining expressions (\ref{Parr}), (\ref{OPE}) and (\ref{Mean2})
gives the desired asymptotic  expression for the scaling functions:
\begin{equation}
\zeta(mr, c(r)) \simeq \sum_{F} A_{F} (mr, c(r))\, (mr)^{\Delta_{F}},
\label{Fin}
\end{equation}
where the summation runs over Galilean invariant scalar operators, with
the coefficient functions $A_{F}$ analytical in their arguments.

Divergences for $mr\to0$ (and hence the anomalous scaling) result from the
contributions of the operators with {\it negative} critical dimensions,
termed ``dangerous'' in \cite{JETP}. Clearly, the leading contribution is
determined by the operator with the lowest (minimal) dimension; the others
determine the corrections. All the operators $\theta^{n}$ are dangerous,
and the spectrum of their dimensions is not restricted from below (there is
no ``most dangerous'' operator); see expression (\ref{KrOp1}). Fortunately,
for {\it a given} correlation function,
only a finite number of those operators
can contribute to the OPE. For (\ref{Parr2}), these are the operators with
$n\le p+k$. Thus,
\begin{equation}
\zeta(mr, c(r)) \simeq \sum_{n=0}^{p+k} A_{n} (mr, c(r))\, (mr)^{\Delta_{n}}
+ \dots
\label{Fin2}
\end{equation}
with $\Delta_{n}$ from (\ref{KrOp1}); the ellipsis stands for the ``more
distant'' corrections, related to the operators with derivatives and other
types of fields. The leading term of the small-$mr$ behavior in (\ref{Fin2})
is given by the operator with the maximum possible $n=p+k$, so that
the final expression has the form
\begin{equation}
\langle \theta^{p}(t,{\bf x}_{1}) \theta^{k}(t,{\bf x}_{2}) \rangle \simeq
\mu^{-(p+k)} (\mu r)^{-\Delta_{p}-\Delta_{k}} (mr)^{\Delta_{p+k}}.
\label{FinF}
\end{equation}

It is worth noting that the set of operators $\theta^{n}$ is ``closed with
respect to the fusion'' in the sense that the leading term in the OPE for
the pair correlator of two such operators is given by the operator from
the same family with the summed exponent. This fact along with the
inequality $\Delta_{p}+\Delta_{k}>\Delta_{p+k}$, which follows from the
explicit expression (\ref{KrOp1}), can be interpreted as the statement
that the correlations of the scalar field in the density model reveal
multifractal behavior; see \cite{DL}.

\subsection{The case of the tracer field} \label{sec:OPET}

For the tracer model, the critical dimensions of the operators $\theta^{n}$
are linear in $n$: $\Delta[\theta^{n}]= n\Delta_{\theta}$, see
eq.~(\ref{KrOp2}). Then the dependence on the separation $r$ in the
asymptotic expressions (\ref{FinF}) disappears: the leading terms of the
inertial-range behavior are constants. More ``vivid'' quantities are
the equal-time structure functions defined as
\begin{widetext}
\begin{equation}
S_{n}(r) = \langle [\theta(t,{\bf x})-\theta(t,{\bf x'}]^{2n}\rangle
= (\nu\mu^{2})^{-n} \eta(\mu r, mr, c/(\mu\nu)),
\quad r=|{\bf x'}={\bf x}|;
\label{struc}
\end{equation}
\end{widetext}
the second equality with dimensionless functions $\eta$ follows from
dimensionality considerations. Solving the RG equations gives the
asymptotic expressions for $\mu r\gg 1$:
\begin{equation}
S_{n}(r) =  (\nu\mu^{2})^{-n} (\mu r)^{-2n\Delta_{\theta}}
\zeta(mr, c(r)),
\label{struc2}
\end{equation}
with $c(r)$ from (\ref{rms}) and some scaling functions $\zeta$.
It is important here that the pair correlation functions
$\langle\theta^{p}\theta^{k}\rangle$ with $k+p=2n$, appearing in the
binomial decomposition of $S_{n}$, have similar
asymptotic representations (\ref{Parr2}) with the same critical dimension
$\Delta_{k}+\Delta_{p}=2n\Delta_{\theta}$, and together they form the single
asymptotic expression (\ref{struc2}). The constant leading terms for
those correlators, related to the contributions of the operator $\theta^{n}$
in the corresponding OPE, cancel each other in the structure function, and
the latter acquires nontrivial dependence on $r$ in the inertial range.

Indeed, both the functions (\ref{struc}) and the action (\ref{Tact}) for the
tracer (not for the density!) are invariant with respect to the constant
shift $\theta(x)\to\theta(x)+{\rm const}$. Then the operators entering the
corresponding OPE,
\begin{widetext}
\begin{equation}
[\theta(t,{\bf x})-\theta(t,{\bf x'}]^{2n} \simeq \sum_{F} C_{F}(mr, c(r))\,
F(t,{\bf x}), \quad r\to 0, \quad {\bf x}=({\bf x}+{\bf x'})/2,
\label{OPE2}
\end{equation}
\end{widetext}
must also be all invariant, so that they can involve the field $\theta$ only
in the form of derivatives. Clearly, the leading term of the small-$m$
behavior will be determined by the scalar operator with maximum possible
number of the fields $\theta$ (namely, $2n$ for the given $S_{n}$) and the
minimum possible number of spatial derivatives (namely, $2n$: one derivative
for each $\theta$). This is nothing other than the operator
$F^{(2n,0)} = (\partial_{i}\theta\partial_{i}\theta)^{n}$ from (\ref{Fnp}).
Thus the desired leading-order expression for $S_{n}$ in the inertial range
is
\begin{equation}
S_{n}(r) \sim  (\nu\mu^{2})^{-n} (\mu r)^{-2n\Delta_{\theta}}
(mr)^{\Delta_{(2n,0)}},
\label{fin2}
\end{equation}
with the dimension $\Delta_{(2n,0)}$ given in (\ref{Dnl}). The operators
$F^{(2p,0)}$ with $p<n$ determine the main corrections to (\ref{fin2}), the
operators with extra derivatives and/or other types of fields correspond
to more ``distant'' corrections (they all must be invariant with respect to
the Galilean transformation and the shift of $\theta$).

For the tracer, the ``multifractal'' behavior is demonstrated by the family
of the operators $F^{(n,0)}$ rather than by the simple powers $\theta^{n}$;
see the end of the preceding subsection. Indeed, it is easy to grasp that
the inertial-range behavior of the pair correlation function
$\langle F^{(p,0)} F^{(k,0)} \rangle$ of two such operators is determined
by the contribution to the OPE from their ``elder brother'' $F^{(n,0)}$
with $n=p+k$ and has the form
(omitting the dependence on the UV parameters $\mu$ and $\nu$)
\begin{equation}
\langle F^{(p,0)}(t,{\bf x}) F^{(k,0)}(t,{\bf x'}) \rangle \sim
r^{-\Delta_{(p,0)}-\Delta_{(k,0)}+\Delta_{(n,0)}}.
\label{MF}
\end{equation}
The required inequality $\Delta_{(n,0)}<\Delta_{(p,0)}+\Delta_{(k,0)}$
\cite{DL} follows from the explicit one-loop expression (\ref{Dnl}).
It remains to note that the operator $F^{(2,0)}$ can be interpreted as
the local dissipation rate of fluctuations of our scalar field.

\subsection{Effects of the large-scale anisotropy} \label{sec:Aniz}

Now consider the effects of the anisotropy, introduced into the system at
large scales $\sim L$ through the correlation function of the random noise
(\ref{noise}). As an illustration, consider first the case of uniaxial
anisotropy: assume that the function $C({\bf r}/L)$ in (\ref{noise})
depends also on a constant unit vector ${\bf n}=\{n_i\}$ that determines
a certain distinguished direction.

Then the irreducible tensor composite operators acquire nonzero mean values,
with the tensor factors built of the vector ${\bf n}$. For example, the mean
value of the operator (\ref{Irr}) is proportional to the irreducible tensor
$n_{i}n_{j}-\delta_{ij}/d$. In general, the mean value of any $l$th rank
irreducible operator is proportional to the tensor
$n_{i_{1}}\dots n_{i_{l}} + \dots$, where the ellipsis stands for the
contributions with the Kronecker $\delta$ symbols that make it irreducible.
Upon substitution into the OPE (\ref{OPE2}), their tensor indices are
contracted with the corresponding indices of the coefficient functions
$C_{F}({\bf r})$. This gives rise to the ($d$-dimensional generalizations
of the) Legendre polynomials $P_{l}(\cos\vartheta)$, where $\vartheta$
is the angle between the vectors ${\bf r}$ and ${\bf n}$.

Thus, the OPE expansion in irreducible composite operators provides the
expansion in the irreducible representations of the $SO(d)$ group.
The main contribution to the ``shell'' with a given $l$ is determined
by the $l$th rank operator with the lowest critical dimension (of course,
it should respect the symmetries of the model and of the left hand side).
Clearly, for the structure function $S_{n}$ and $l \le 2n$ the needed
operator is $F^{(2n,l)}_{i_{1}\dots i_{l}}$ from (\ref{Fnp}). For $l>2n$
we need the operators that contain more derivatives than fields.

The expansion that takes into account only the leading term in each
shell has the form (again, we omit $\nu$ and $\mu$):
\begin{equation}
S_{n} = r^{-2n\Delta_{\theta}} \sum_{l=0}^{2n} A_{l}(r)\,
P_{l}(\cos\vartheta)\, (mr)^{\Delta_{(2n,l)}} + \dots
\label{shells}
\end{equation}
with the dimension $\Delta_{(2n,l)}$ from (\ref{Krnl}); the ellipsis stands
for the contributions with $l>2n$. For the general large-scale anisotropy,
all the spherical harmonics $Y_{l\!s}$ will appear in the expansion, with
the exponents depending only on $l$.

From the explicit leading-order expressions (\ref{Dnl}) it follows that the
dimensions (\ref{Krnl}), for a fixed $n$, monotonically increase with $l$:
\begin{equation}
\Delta_{n,l}>\Delta_{n,p}\quad {\rm if}\ l>p,
\label{hier}
\end{equation}
or, in the differential form, $\partial\Delta_{n,l}/\partial l>0$. Similar
inequalities were derived earlier in various models of passively advected
vector \cite{Lanotte} and scalar \cite{A99} fields. This fact has
a clear physical interpretation: in the presence of the large-scale
anisotropy, anisotropic contributions in the inertial range exhibit
an hierarchy, related to the ``degree of anisotropy'' $l$: the leading
contribution is given by the isotropic shell ($l=0$); the corresponding
anomalous exponent is the same as for the purely isotropic case. The
contributions with $l>1$ give only corrections which become relatively
weaker as $mr\to0$, the faster the higher the degree of anisotropy $l$
is. This effect gives quantitative support for Kolmogorov's hypothesis
of the local isotropy restoration and appears rather robust, being
observed for the real fluid turbulence \cite{Hier}.

The hierarchy (\ref{hier}) becomes more strongly pronounced as the degree of
compressibility $\alpha$ increases, which can be expressed by the inequality
$\partial^{2}\Delta_{n,l}/\partial l\partial {\alpha}>0$. Thus the
anisotropic corrections become further from one another and from the
isotropic term, in contrast to the situation observed earlier for passive
scalar \cite{AKens,AK2} and vector \cite{alpha} fields, advected by
Kraichnan's ensemble. The same inequality holds for the ``frozen'' regime
in the Gaussian model with finite correlation time, the fact overlooked
in \cite{AKens}.

For $l>2n$, the leading contributions to the $l$th shell are determined by
the operators that involve more derivatives than fields. The calculation
of their dimensions is a difficult task because of the mixing of such
operators in renormalization. The hierarchy relations remain valid due to
the contributions of the canonical dimensions to the general expression
(\ref{Krit}): clearly, their critical dimensions have the forms $l-2n+O(y)$.

Fortunately, for the pair correlation functions, the full analog of the
expression (\ref{shells}) can be presented, with all the shells included.
Indeed, it is clear that the
leading term of the $l$th shell now is determined by the single operator
(\ref{Fl}) with two fields and $l$ tensor indices: it is unique up to
derivatives, which have vanishing mean values and do not contribute to the
quantities of interest. Thus the desired asymptotic expression has the form
\begin{equation}
\langle\theta(t,{\bf x}) \theta(t,{\bf x}') \rangle = r^{-2\Delta_{\theta}}
\sum_{l=0}^{\infty} A_{l}(r)\,
P_{l}(\cos\vartheta)\, (mr)^{\Delta_{l}}
\label{Orujo}
\end{equation}
with the dimensions $\Delta_{l}$ from (\ref{Dl}), (\ref{gamZ}) for the
tracer and (\ref{Dl}), (\ref{ana7}) for the density case. The hierarchy
of anisotropic contributions, similar to (\ref{hier}), holds, at least
for small $y$, due to the contribution of the canonical dimensions to
(\ref{Dl}): $\Delta_{l}=l-2+O(y)$. Thus the leading term in (\ref{Orujo})
is given by the scalar operator $\theta^{2}$. When one passes to the
structure function $S_{2}$ for the tracer that term is subtracted, and the
leading role is inherited by the scalar operator $F^{(2,0)}$ from (\ref{Fnp})
in agreement with (\ref{shells}). The hierarchy is getting weaker as the
compressibility parameter $\alpha$ grows:
$\partial^{2}\Delta_{n,l}/\partial l\partial {\alpha}<0$, as follows
from the analysis of the explicit one-loop expressions (\ref{Dl}),
(\ref{gamZ}), (\ref{ana7}). Here our results agree with those for
the Kraichnan model: anisotropic corrections become closer
to each other and to the isotropic term; cf.~\cite{AG}.

\section{Discussion and conclusion} \label{sec:DC14}

We have studied two models of passive scalar advection: the case of the
density of a conserved quantity and the case of a tracer, described by the
advection-diffusion equations (\ref{density1}) and (\ref{tracer1}),
respectively, and subject to a random large-scale forcing (\ref{noise}).
The advecting velocity field is described by the Navier--Stokes equations
for a compressible fluid (\ref{ANU}), (\ref{ANU1}) with
an external stirring force with the correlation function
$\propto k^{4-d-y}$; see (\ref{force}), (\ref{power}).

The full stochastic problems can be formulated as field theoretic models
with the action functionals specified in (\ref{action}), (\ref{Dact}) and
(\ref{Tact}). Those models appear multiplicatively renormalizable, so that
the corresponding RG equations can be derived in a standard fashion.
They have the only IR attractive fixed point in the physical range of the
model parameters, and the correlation functions reveal scaling behavior in
the IR region (inertial and energy-containing ranges).

Their inertial-range behavior was studied by means of the OPE; existence
of anomalous scaling (singular power-like dependence on the integral scale
$L$) was established. The corresponding anomalous exponents were identified
with the scaling (critical) dimensions of certain composite fields
(composite operators): powers of the scalar field for the density and
powers of its spatial gradients for the tracer, so that they can be
systematically calculated as series in the exponent $y$. The practical
calculations were performed in the leading order (one-loop approximation)
and are presented in (\ref{KrOp1}), (\ref{Dnl}). The results
(\ref{KritEx}), (\ref{KriTet}) for primary fields and (\ref{KrOp2})
for the operators $\theta^{n}$ for the tracer are given by this
approximation exactly.

Thus we removed two important restrictions of the previous treatments of the
passive compressible problem: absence of time correlations and Gaussianity
of the advecting velocity field. We stress that in contrast to previous
studies that combined compressibility with finite correlation time
\cite{AKens,AK2}, the present model is manifestly Galilean covariant,
and this fact holds in all orders of the perturbation theory.

In a few respects, however, the results obtained here are very similar
to those obtained earlier for the compressible version of Kraichnan's
rapid-change model \cite{RG1,AH,AG} and the Gaussian model with finite
correlation time \cite{AKens,AK2}. First of all, the mechanism of the origin
of anomalous scaling is essentially the same: the anomalous exponents are
identified with the dimensions of {\it individual} composite operators.

Second, those dimensions are insensitive to the specific choice of the
random force (\ref{noise}), because the propagator
$\langle\theta\theta\rangle_{0}$ does not enter into the relevant Green
functions. (In particular, this means that the anomalous exponents
remain intact if the artificial noise is replaced by an imposed
linear gradient, a more realistic formulation of the problem).
The force maintains the steady state and thus provides nonvanishing
mean values for the composite operators, but it does not affect their
dimensions.

For the rapid-change case, this fact is naturally interpreted within the
zero-mode approach, where the equal-time correlation functions satisfy
certain differential equations, and the anomalous exponents are related
to the solutions of their {\it homogeneous} analogs, where the forcing
terms are discarded; see \cite{FGV,GK}. On the contrary, the amplitudes
are found by matching of these inertial-range zero-mode solutions with
the large-scale solutions of the full inhomogeneous equations,
which are nontrivial only in the presence of the forcing terms.

The close resemblance in the RG+OPE pictures of the origin of anomalous
scaling for the present model and its rapid-change predecessors suggests
that for the former, the concept of zero modes (and thus that of statistical
conservation laws) is also applicable, although no closed differential
equations can be derived for the equal-time correlation functions.

Although the anomalous exponents are independent of the specific choice of
the noise, they do depend on the exponent $y$, the dimension of space $d$,
and on the parameter $\alpha$ that measures the degree of compressibility.
In this respect, our results are also similar to those obtained for simpler
models. An important difference with Gaussian models appears when possible
dependence on the time scales is studied. It was argued that the exponents
can depend on more details of the velocity ensemble than only the exponents,
namely, on the dimensionless ratio of the correlation times of the scalar
and velocity fields; see {\it e.g.} the discussion in~\cite{ShS}. Indeed,
analytic results obtained for Gaussian models with a finite correlation
time within the zero-mode technique \cite{Falk3} and the RG+OPE approach
\cite{A99,AKens,AK2} show that such a dependence indeed takes place, at
least for some of the possible scaling regimes.

In the present case, the exponents could depend, in principle, on the
dimensionless parameters $u_{0}$, $v_{0}$, $w_{0}$, the ratios of various
viscosity and diffusion coefficients. After the RG treatment, these
parameters are replaced with the corresponding invariant variables,
which exactly have the meaning of the ratios of the correlation times
of the transverse and longitudinal components of the velocity field, the
pressure and the scalar field; for a detailed discussion of this issue
see~\cite{A99}. Existence of the unique IR attractive fixed point shows
that in the IR range these ratios tend to their fixed-point values $u_{*}$,
$v_{*}$, $w_{*}$ irrespective of the initial values $u_{0}$ {\it etc.}
We conclude that the anomalous exponents are independent on the time scale;
the dependence observed in previous treatments is an artifact of simplified
Gaussian statistics.

Another essential difference between our results and those obtained
for Kraichnan's rapid-change model is that in the latter the anomalous
exponents have a finite limit when the parameter that measures
degree of compressibility (analog of $\alpha$ from (\ref{power}) in our
model) goes to infinity, that is, for the purely
potential velocity field. In our case all the nontrivial anomalous
dimensions grow with $\alpha$ without bound. Formally, the difference
is due to the fact
that the coordinate of the fixed point (\ref{FP}) in our model is
independent on $\alpha$ and the dependence on it appears only in the
numerators of the expressions like (\ref{KrOp1}), (\ref{Dnl}). This fact
also means that the one-loop contributions in the critical dimensions
become large as $\alpha$ grows,
and the one-loop approximation can hardly be trusted even for small $y$.
One may think that the real RG expansion parameter then becomes
$y\alpha$ rather than $y$.
In this connection we also recall that the IR attractive fixed point in the
one-loop approximation exists for all $\alpha$ but
ceases to exist for $\alpha=\infty$ (purely potential forcing).
These facts suggest that, beyond
the one-loop approximation, the fixed point (\ref{FP}) in fact disappears
or loses its stability, and the corresponding scaling regime
 undergoes some qualitative changeover, the possibility
supported by the phase transition to a purely chaotic state
observed in \cite{tracer} for a simplified model.

To investigate this issue, it is necessary to go beyond the
leading one-loop approximation (of course, starting with the compressible
Navier--Stokes equation itself) and to discuss the existence, stability and
the dependence on $\alpha$ of the fixed point at least at the two-loop
level, which seems to be a difficult technical task. Another interesting
generalization of our present investigation is to derive a more realistic
expression for the random force correlator (\ref{power}) in order to
determine realistic values of $\alpha$ and to express it in terms of
measurable quantities. Here the combination of the
RG techniques and the energy balance equation seems promising; see
\cite{spectra} for the incompressible case. This work remains for the
future and is partly in progress.

\section*{Acknowledgments}

The authors are indebted to L.Ts. Adzhemyan, Michal Hnatich, Juha Honkonen
and M.Yu. Nalimov for discussion.
The authors acknowledge Saint Petersburg State University for
the research grant 11.38.185.2014.
The work was also supported by the Russian Foundation for Basic
Research within the project~12-02-00874-a.



\end{document}